\documentclass[twocolumn,article,amsmath,amssymb,aps,prd,floatfix,showkeys,superscriptaddress]{revtex4-1}
\usepackage[utf8x]{inputenc} 
\usepackage{bm}
\usepackage{amssymb}
\usepackage{amsmath}
\usepackage{graphicx}
\usepackage{epsfig}
\usepackage{array}
\usepackage{float}
\usepackage{color}
\usepackage[usenames,dvipsnames]{xcolor}
\usepackage{epstopdf}
\usepackage{natbib}
\usepackage{lineno}
\usepackage{slashed}
\usepackage{color}
\usepackage{caption}
\usepackage{hyperref}
\usepackage{slashed}
\usepackage{soul}
\usepackage[bottom]{footmisc}
\newcommand{\T}{\tilde}

\newcommand{\be}{\begin{equation}}
\newcommand{\ee}{\end{equation}}
\newcommand{\bea}{\begin{eqnarray}}
\newcommand{\eea}{\end{eqnarray}}
\newcommand{\beas}{\begin{eqnarray*}}
\newcommand{\eeas}{\end{eqnarray*}}

\newcommand{\slsh}[1]{{\not \! #1}}

\newcommand{\dpi}{(2\pi)}
\newcommand{\p}{\parallel}
\newcommand{\pp}{\perp}
\newcommand{\ga}{\gamma_{\alpha}}
\newcommand{\gm}{\gamma_{\mu}}
\newcommand{\gn}{\gamma_{\nu}}

\newcommand{\ps}{\slashed{p}}

\newcommand{\gmn}{g^{\mu\nu}}
\newcommand{\gma}{g^{\mu\alpha}}
\newcommand{\gan}{g^{\alpha\nu}}

\newcommand{\nn}{\nonumber}
\newcommand{\eB}{\left|eq_f B\right|}
\newcommand{\wwp}{\omega_p}
\newcommand{\wq}{\omega_q}

\begin{document}
\title{Centrality dependence of photon yield and elliptic flow from gluon fusion and splitting induced by magnetic fields in relativistic heavy-ion collisions}
\author{Alejandro Ayala}
\affiliation{Instituto de Ciencias
  Nucleares, Universidad Nacional Aut\'onoma de M\'exico, Apartado
  Postal 70-543, CdMx 04510,
  Mexico.}
 \affiliation{Centre for Theoretical and Mathematical Physics, and Department of Physics, University of Cape Town, Rondebosch 7700, South Africa.}
\author{Jorge David Casta\~no-Yepes}
\email{Corresponding author.\\
E-mail address: jorgecastanoy@gmail.com (J.D. Casta\~no-Yepes).}
\affiliation{Instituto de Ciencias
  Nucleares, Universidad Nacional Aut\'onoma de M\'exico, Apartado
  Postal 70-543, CdMx 04510,
  Mexico.}
\author{Isabel Dominguez Jimenez}
\affiliation{Universidad Aut\'onoma de Sinaloa,
Avenida de las Am\'ericas y Boulevard Universitarios, Ciudad Universitaria,
C.P. 80000, Culiac\'an, Sinaloa, M\'exico.}
\author{Jordi Salinas San Mart\'in}
\affiliation{Instituto de Ciencias
  Nucleares, Universidad Nacional Aut\'onoma de M\'exico, Apartado
  Postal 70-543, CdMx 04510,
  Mexico.}
\author{Mar\'ia Elena Tejeda-Yeomans}
\affiliation{Facultad de Ciencias - CUICBAS, Universidad de Colima, Bernal D\'iaz del Castillo No. 340, Col. Villas San Sebasti\'an, 28045 Colima, Mexico.}
%
%
\begin{abstract}
We compute the photon yield and elliptic flow coefficient in relativistic heavy-ion collisions from gluon fusion and splitting processes induced by a magnetic field  for different centralities. The calculation accounts for the intense magnetic field and the high gluon occupation number at early times. The  photon production induced by these process represents an excess contribution over calculations without magnetic field effects. We compare this excess to the difference between PHENIX data and recent hydrodynamic calculations for the photon transverse momentum distribution and elliptic flow coefficient $v_2$. The time evolution of the field strength and reaction volume is computed using UrQMD. We show that with reasonable values for the saturation scale, the calculation helps to better describe the experimental results obtained at RHIC energies for the lowest part of the transverse photon momentum at different centralities.
\end{abstract}  
%
%
\maketitle

\section{Introduction}\label{I}

The possibility to produce large strength magnetic fields in peripheral heavy-ion collisions~\cite{intensity} serves also as a motivation to explore new channels~\cite{Skokov,Zakharov,Tuchin} to try explain the anomalous excess of direct photons produced in these reactions, the so called {\it direct photon puzzle}~\cite{experimentsyield,experimentsv2,experiments2}. Since a magnetic field induces the breaking of rotational invariance, this field is not only a source for an excess photon yield but also of an increase of the second harmonic coefficient ($v_2$) of the Fourier expansion in the azimuthal photon  distribution~\cite{v2photons}. Although some recently improved hydrodynamic~\cite{hydro-photons1,hydro-photons2}
and transport~\cite{transport} calculations obtain a better agreement with ALICE and PHENIX measurements of low and intermediate transverse momentum ($p_T$) photons, this agreement is not yet complete~\cite{review}. Moreover, PHENIX~\cite{phenix-soft-photons-2018} has also found that in Au+Au and Cu+Cu collisions at different centralities and beam energies, the yield of low $p_T$ photons ($\lesssim 2$ GeV) scales with a given power of $N_{\mathrm{coll}}$, which suggests that the source of these photons is very similar across beam energies and colliding species. In a recent work~\cite{our-old-v2} we have shown that an important source of low $p_T$ photons is provided by the gluon fusion induced by the presence of a magnetic field during the early stages of a high-energy heavy-ion collision. This process also increases $v_2$ at low $p_T$. The study was performed resorting to an educated guess for the intensity of the magnetic field and with an estimate of the space-time volume where the reaction occurs. We compared our results for the excess photon yield and $v_2$ with PHENIX data subtracting the state of the art computation of Ref.~\cite{hydro-photons1} for 20-40\% Au+Au collisions at RHIC and found a reasonable agreement for low $p_T$ photons. In this work we use instead Ultrarelativistic Quantum Molecular Dynamics (UrQMD)~\cite{UrQMD} to  estimate an upper limit for the time evolution of the magnetic field intensity and the volume produced by the collision participants and spectators. This is achieved producing samples of Au+Au and Cu+Cu collisions at $\sqrt{s_{\mathrm{NN}}}=200$ GeV for different centrality ranges. In our approach, the influence of medium effects for the magnetic field evolution are not accounted for. Estimates of the medium influence, encoded in terms of a finite conductivity, permittivity and permeability, are discussed for instance in Refs.~\cite{Tuchin1,Tuchin2,Gursoy,Li,Inghirami}. We also extend the analysis of our previous work to include the contribution to photon production from another closely related process to gluon fusion, namely, gluon splitting into a photon and a gluon, to compute both the photon yield and $v_2$. We use this approach to perform a centrality dependence study for the excess photon yield and $v_2$ comparing to the recent PHENIX measurements. We find a relatively good agreement for the lower part of the spectra which improves for peripheral collisions and when the magnetic field strength includes the contribution from both spectators and participants.  Another approach  to include the effects of electromagnetic fields for photon production in heavy-ion collisions is worked in Ref.~\cite{Deng}, where
the photon production cross section is estimated from proton-neutron bremsstrahlung in a Boltzmann-Uehling-Uhlenbeck model, accounting for the
electromagnetic field in the simulated reactions.

Our manuscript is organized as follows: In Sec.~\ref{2} we revisit the field theoretical calculation of the gluon fusion process in the presence of a constant magnetic field. We also include the contribution to photon production from gluon splitting. In Sec.~\ref{III} we use UrQMD to compute the time evolution of the magnetic field strength and reaction volume that are needed for a more reliable computation of the photon yield and second harmonic coefficient. In Sec.~\ref{IV} we present the results for the photon yield and $v_2$ and compare to PHENIX experimental data. We finally summarize and conclude in Sec.~\ref{V}.

\section{Soft-photon yield and second harmonic coefficient}\label{2}

In order to present the calculation in a self-contained manner, here we summarize the field theoretical framework to obtain the photon yield and $v_2$. More details can be found in Ref.~\cite{our-old-v2}. Since the presence of a magnetic field breaks translational invariance, the amplitude for the process has to be computed in coordinate space and subsequently integrated over space-time. 

\begin{figure}
\begin{center}
\includegraphics[scale=.47]{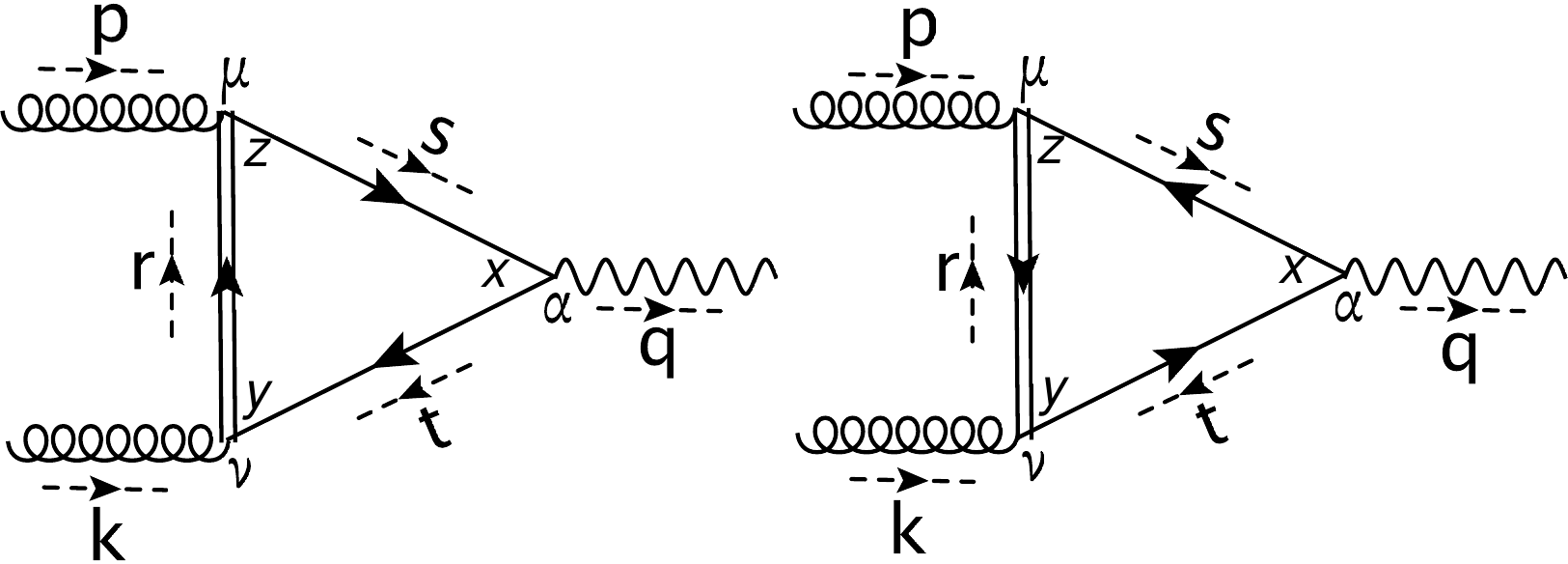}\\
\includegraphics[scale=.47]{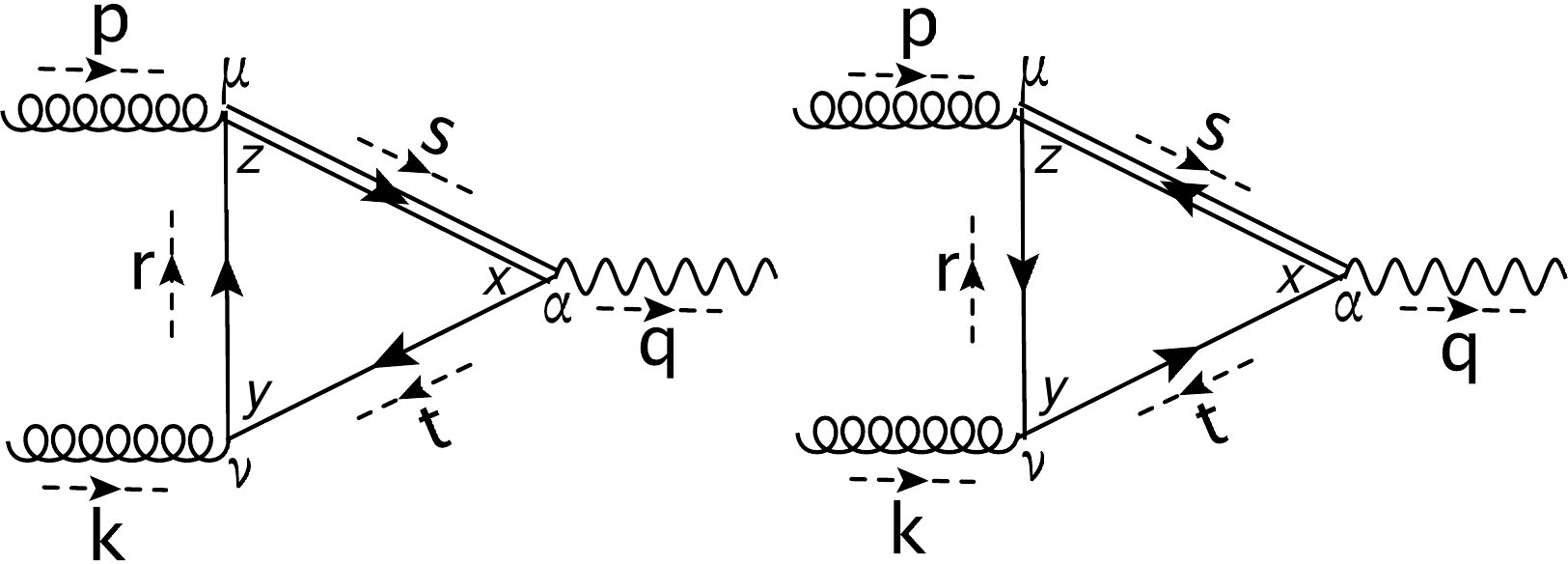}\\
\includegraphics[scale=.47]{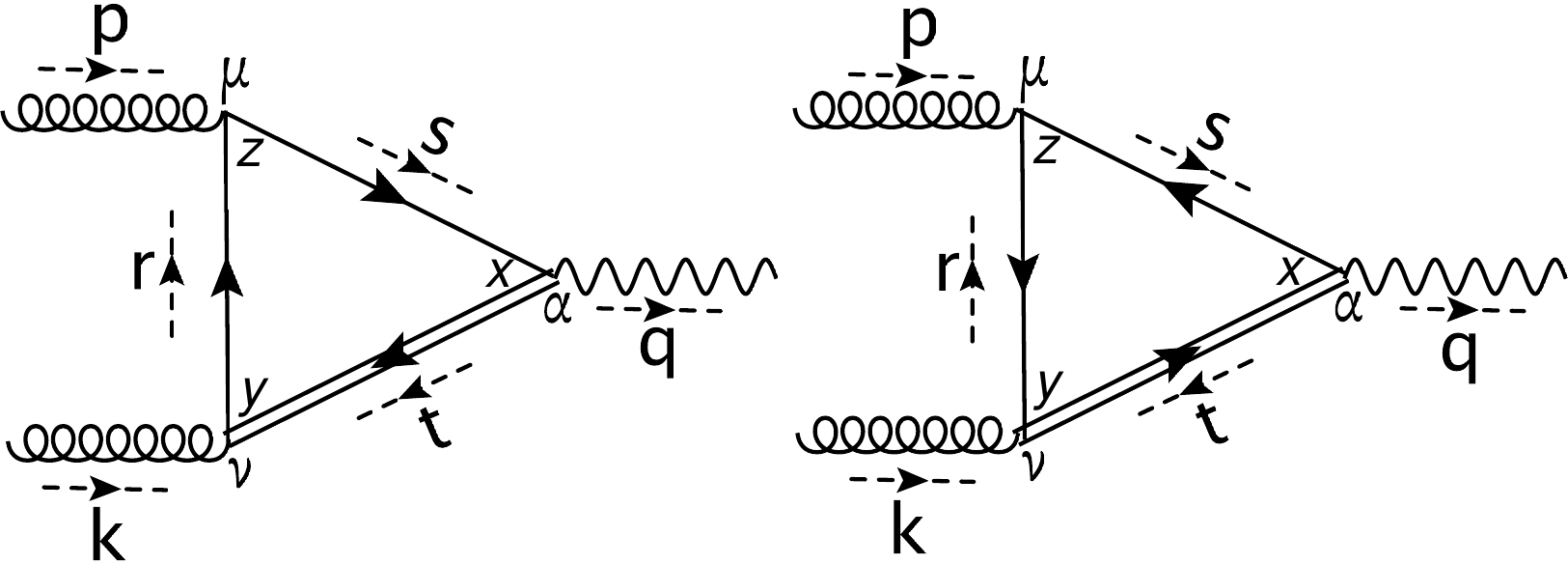}
\caption{Leading order contribution for photon production from gluon fusion in the presence of a magnetic field where the double lines represent that the corresponding quark propagator is in the 1LL, whereas the single lines represent the propagator in the LLL. The arrows in the propagators represent the charge flow and the arrows on the sides represent the momentum direction.}
\label{Diag1}
\end{center}
\end{figure}

\begin{figure}
\begin{center}
\includegraphics[scale=.49]{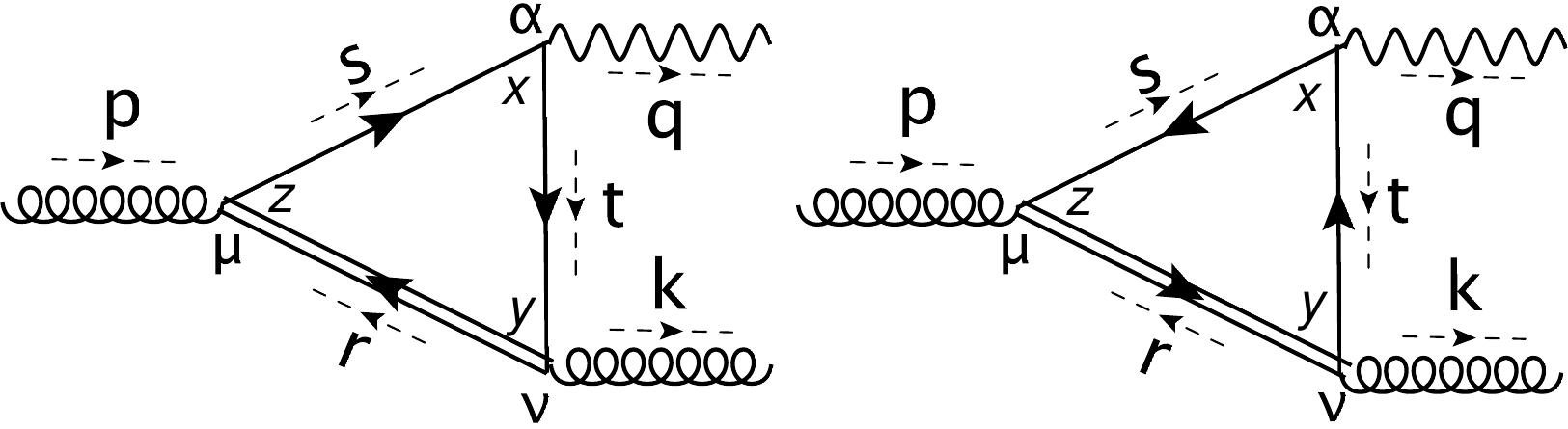}\\
\includegraphics[scale=.49]{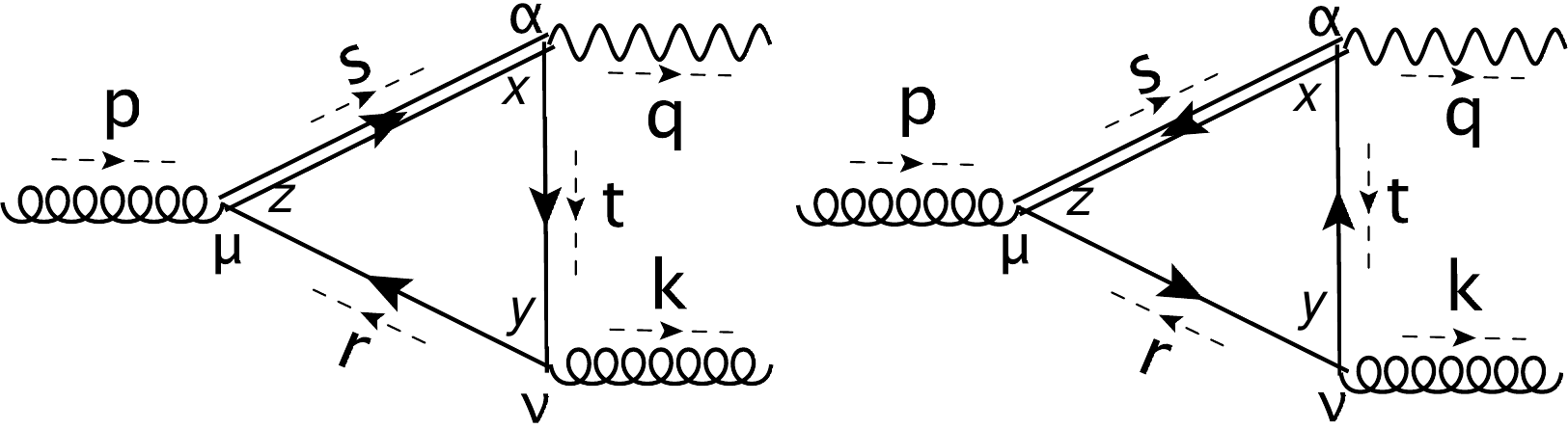}\\
\includegraphics[scale=.49]{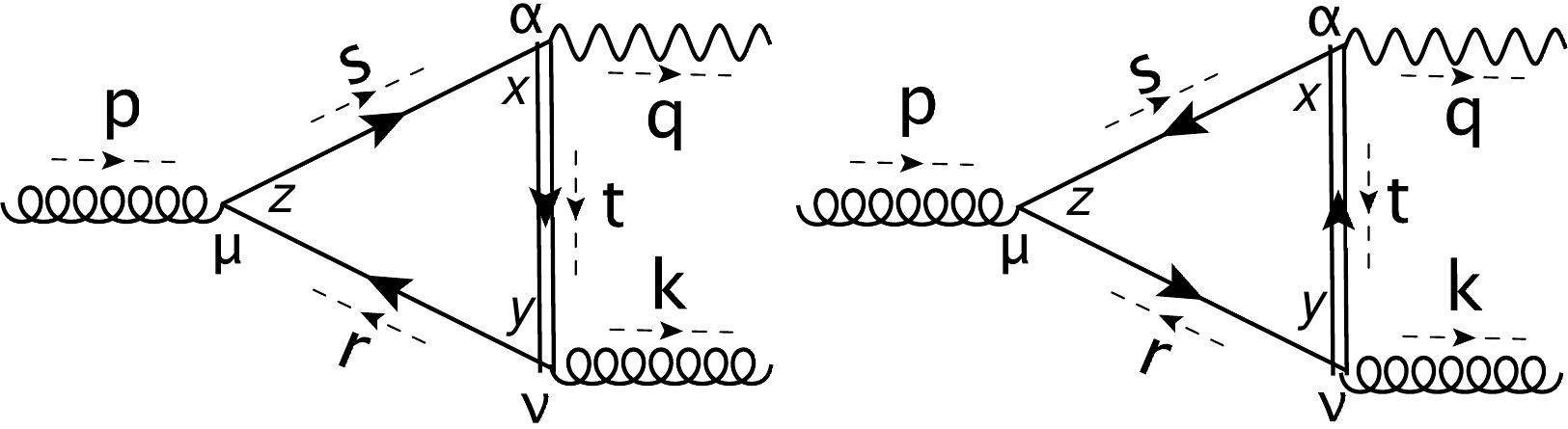}
\caption{Leading order contribution for photon production from gluon splitting in the presence of a magnetic field where the double lines represent that the corresponding quark propagator is in the 1LL, whereas the single lines represent the propagator in the LLL. The arrows in the propagators represent the charge flow and the arrows on the sides represent the momentum direction.}
\label{Diag2}
\end{center}
\end{figure}

The lowest order process in the strong, $\alpha_\text{s}=g^2/4\pi$, and electromagnetic, $\alpha_{\tiny{\text{em}}}=e^2/4\pi$, couplings come from amplitudes made of quark {\it triangle} diagrams with two gluons and one photon attached each at one of the vertices of the triangle. In this work we include both the gluon fusion and gluon radiation channels, depicted in Figs.~\ref{Diag1} and~\ref{Diag2}, respectively, where each row of diagrams corresponds to the two possible ways the charge flows in a given triangle.


The coordinate representation of the quark propagator is given by~\cite{Schwinger}
\bea
S(x,x')=\Phi(x,x')\int\frac{d^4p}{\dpi^4}e^{-p\cdot(x-x')}S(p),
\label{propagator}
\eea
where $S(p)$ is the Fourier transform to momentum space of the translational invariant part of the propagator, i.e.

\bea
i S(p)&=&\int _{0}^{\infty}\frac{d\tau}{\cos(|eq_fB|\tau)}e^{i\tau\left[p_\p^2-p_\pp^2\frac{\tan\left(\eB\tau\right)}{\eB\tau}-m_f^2+i\epsilon\right]}\nn\\
&\times&\Bigg{\{}\left[\cos(\eB\tau)+\gamma_1\gamma_2\sin(\eB\tau)\right](m_f+\slashed{p}_\p)\nn\\
&-&\frac{\slashed{p}_\pp}{\cos(\eB\tau)}\Bigg{\}},
\eea
and $\Phi(x,x')$ is the phase factor which carries the explicit coordinate dependence, and is given by
\bea
\Phi(x,x')=e^{ieq_f\int_{x'}^xd\xi^\mu\left[A_\mu+\frac{1}{2}F_{\mu\nu}(\xi-x')^\nu\right]},
\label{phasefactor}
\eea
with $eq_f$ the quark charge in units of $e$, the absolute value of the electron charge.

In this section, we study the case of a constant magnetic field chosen to point in the $z$-direction. In the next section we will use this setup to study the case where the magnetic field evolves with time. A constant magnetic field in the $z$-direction is obtained from a vector potential $A^\mu$ in the symmetric gauge $A^\mu=\frac{B}{2}(0,-y,x,0)$. Also, for a four-momentum $p^\mu$, we have defined $p_\pp^\mu=(0,p_1,p_2,0)$, $p_\p=(p_0,0,0,p_3)$, $p_\pp^2=p_1^2+p_2^2$, $p_\p^2=p_0^2-p_3^2$, and therefore $p^2=p_\p^2-p_\pp^2$.

From the Feynman diagrams of Fig.~\ref{Diag1}, we can write the expression for the amplitude $\widetilde{{\mathcal{M}}}_{gg\rightarrow\gamma}$, and this is explicitly given by
\bea
   \widetilde{{\mathcal{M}}}_{gg\rightarrow\gamma}&=&-\int \!d^4xd^4yd^4z\int\!\frac{d^4r}{(2\pi)^4}
   \frac{d^4s}{(2\pi)^4}\frac{d^4t}{(2\pi)^4}\nn\\
   &\times&e^{-it\cdot (y-x)}e^{-is\cdot (x-z)}e^{-ir\cdot (z-y)}e^{-ip\cdot z}e^{-ik\cdot y}e^{iq\cdot x}\nonumber\\
   &\times&
   \Big\{
   {\mbox{Tr}}\left[ ieq_f\gamma_\alpha iS_{ac}(s) ig\gamma_\mu t^c iS_{cd}(r) ig\gamma_\nu t^d iS_{da}(t) \right]\nn\\
   &+&
   {\mbox{Tr}}\left[ ieq_f\gamma_\alpha iS_{ad}(t) ig\gamma_\nu t^d iS_{dc}(r) ig\gamma_\mu t^c iS_{ca}(s) \right]
   \Big\}
   \nonumber\\
   &\times&\Phi(x,y)\Phi(y,z)\Phi(z,x)\epsilon^\mu(\lambda_p)\epsilon^\nu(\lambda_k)\epsilon^{\alpha}(\lambda_q),
   \label{amplitude}
\eea
where $p$ and $k$ are the gluon and $q$ the photon four-mo\-men\-ta, $t^c, t^d$ are Gell-Mann matrices, and the polarization vectors for the gluons and the photon are $\epsilon^\mu(\lambda_p)$, $\epsilon^\nu(\lambda_k)$, $\epsilon^\alpha(\lambda_q)$, respectively. The Lorentz indices $\mu, \nu,\alpha$ and the space-time coordinates $x,y,z$ associated to each vertex, are also depicted in Fig.~\ref{Diag1}. The product of phase factors is
$\Phi(x,y)\Phi(y,z)\Phi(z,x)=\exp\left\{i\frac{\eB}{2}\epsilon_{mj}(z-x)_m(x-y)_j\right\}$,
where the indices $m,j=1,2$ and $\epsilon_{mj}$ is the Levi-Civita tensor. 

In order to simplify the calculation, we use the fact that at the earliest times, when thermalization has not been achieved, the magnetic field is the dominant internal energy-scale in the process. Thus the calculation accounts for photons coming from the shattered {\it glasma}, where gluons are described by a dense, non-equilibrium, state~\cite{Larry2}. Therefore, in the absence of thermal effects, and with $|eq_fB|>>m_f^2$, one could in principle work using the quark propagators in the lowest Landau level (LLL). However, it can be shown that if all the quarks in the triangle are in the LLL, the amplitude vanishes. The leading order contribution is obtained when at least one of the quarks in the loop is in the first excited Landau level (1LL). This is depicted in Figs.~\ref{Diag1} and \ref{Diag2} where the single line in the loop represents a quark in the LLL whereas the double line represents a quark in the 1LL. The corresponding expressions for the quark propagators, in the massless limit ($m_f \to 0$), are
\bea
iS^{(0)}_{ab}(p)&=&i\frac{e^{-p_\pp^2/eq_fB}}{p_\p^2}\delta_{ab}\ps_\p\mathcal{O}^{-}
\nonumber
\eea
\bea
iS^{(1)}_{ab}(p)&=&-2i\frac{e^{-p_\pp^2/eq_fB}}{p_\p^2-2\left |eq_fB\right |}\delta_{ab}\nn\\
&\times&\left[\ps_\p\mathcal{O}^{-}\left(1-\frac{2p_\pp^2}{\left | eq_fB \right |}\right)-\ps_\p\mathcal{O}^{+}+2\ps_\pp\right]\!,
\label{propagadores2}
\eea
where $a,\ b$ are color indices and the operators 
\bea
\mathcal{O}^{(\pm)}=\left[1\pm i\gamma^1\gamma^2\text{sign}(eq_f B)\right]/2
\nonumber
\eea
are projectors onto
the longitudinal space with $\text{sign}(eq_f B)$ representing the sign of the charge flowing in the quark line. With this approximation, the matrix element (omitting color indices, for simplicity) in Eq. (\ref{amplitude}) becomes 
\bea
&&\widetilde{{\mathcal{M}}}_{gg\rightarrow\gamma}=8i\dpi^4\delta^{(4)}\left(q-k-p\right)
\delta^{cd}eq_f g^2\nonumber\\
&\times&\int\frac{d^4r}{\dpi^4}\frac{d^4s}{\dpi^4}\frac{d^4t }{\dpi^4}\;\epsilon^{\mu}(\lambda_p)\epsilon^\nu(\lambda_k)\epsilon^{\alpha}(\lambda_q)\nonumber\\
&\times&\int d^4w\;d^4l\, e^{-il(r-t-k)}e^{-iw(r-s+p)}\nn\\
&\times&\exp\left\{-i\frac{|eq_fB|}{2}\epsilon_{mj}w_ml_j\right\}
\exp\left\{-\frac{r_\pp^2+s_\pp^2+t_\pp^2}{|eq_fB|}\right\}\nonumber\\
&\times&\text{Tr}\left\{\frac{\gamma_1\gamma_2\ga\slsh{t}_\pp\gn\slsh{r}_\p\gm^{\p}\slsh{s}_\p}{r_\p^2s_\p^2\left(t_\p^2-2\left| eq_fB\right|\right)}
+\frac{\gamma_1\gamma_2\gm\slsh{s}_\pp\ga\slsh{t}_\p\gn^{\p}\slsh{r}_\p}{t_\p^2r_\p^2\left(s_\p^2-2\left| eq_fB\right|\right)}\right.\nn\\
&+&\left.\frac{\gamma_1\gamma_2\gn\slsh{r}_\pp\gm\slsh{s}_\p\ga^{\p}\slsh{t}_\p}{s_\p^2t_\p^2\left(r_\p^2-2\left| eq_fB\right|\right)}\right\}\epsilon_\mu(\lambda_p)\epsilon_\nu(\lambda_k)\epsilon_\alpha(\lambda_q).
\label{matrixelem}
\eea

The integral in Eq.~(\ref{matrixelem}) has a complicated structure. A considerable simplification can be attained in two limiting cases: either ignoring (a) the loop momentum that is being compared to $|eq_fB|$ in the denominators or (b) the magnetic field strength in the denominators compared to the loop momenta. In this work we focus in the low photon momentum part of the spectrum and thus will consider case (a), namely, that $|eq_fB|$ is large with respect to the loop momenta. Since, through momentum conservation in the vertices, a small loop momenta is tantamount of small external momenta, the result will be valid only for low photon energies. The more general case is currently being explored and will be reported elsewhere. In this spirit, for
central rapidity and for gluons in the shattered glasma with momenta much less than the saturation scale ($\Lambda_s$), it is possible to further simplify the denominators of Eq.~(\ref{matrixelem}) considering that $2|eq_fB|\gg t_\p^2,\ s_\p^2,\ r_\p^2$. After a straightforward calculation, the matrix element can be written as
\bea
&&\widetilde{\mathcal{M}}_{gg\rightarrow\gamma}=-i\dpi^4\delta^{(4)}(q-k-p)\frac{eq_fg^2\delta^{cd}e^{f(p_\pp,k_\pp)}}{32\pi\dpi^8}\nn\\
&\times&\left\{\left(\gma_\p-\frac{p^\mu_\p p^\alpha_\p}{p_\p^2}\right)h^\nu(a)-\left(\gmn_\p-\frac{p^\mu_\p p^\nu_\p}{p_\p^2}\right)h^\alpha(a)\right.\nn\\
&+&\left.\left(\gmn_\p-\frac{k^\mu_\p k^\nu_\p}{k_\p^2}\right)h^\alpha(b)-\left(\gan_\p-\frac{k^\alpha_\p k^\nu_\p}{k_\p^2}\right)h^\mu(b)\right.\nn\\
&+&\left.\left(\gan_\p-\frac{q^\alpha_\p q^\nu_\p}{q_\p^2}\right)h^\mu(c)-\left(\gma_\p-\frac{q^\mu_\p q^\alpha_\p}{q_\p^2}\right)h^\nu(c)\right\}\nn\\
&\times&\epsilon_\mu(\lambda_p)\epsilon_\nu(\lambda_k)\epsilon_\alpha(\lambda_q),
\label{matrint}
\eea
where the trace over the Gell-Mann matrices has also been performed and $h^{\mu}(a)=-(i/\pi)\epsilon_{ij}a^ig^{j\mu}_\pp$, $a_i=p_i + 2k_i + i\epsilon_{im}p_m$, $b_i=2p_i + k_i - i\epsilon_{im}k_m$, $c_i=k_i - p_i + i\epsilon_{im}(p_m + k_m)$, with
\bea
f\left(p_\pp,k_\pp\right)&=&\frac{1}{8|eq_fB|}\left(p_m-k_m+i\epsilon_{mj}(p_j+k_j)\right)^2\nonumber\\
&-&\frac{1}{2|eq_fB|}\left(p_m^2+k_m^2+2i\epsilon_{jm}p_mk_j\right),
\eea
where $g_\pp={\mbox{diag}}(1,1)$ and $g_\parallel={\mbox{diag}}(1,-1)$ are the metric tensors in the transverse and longitudinal spaces.

The diagrams in Fig.~\ref{Diag2} contribute to the matrix element for gluon splitting, $\mathcal{\T{M}}_{g\rightarrow g\gamma}$. This process is related to the gluon fusion channel, $\mathcal{\T{M}}_{gg\rightarrow\gamma}$, by the crossing symmetry
\bea
\widetilde{\mathcal{M}}_{g\rightarrow g\gamma}(p,k,q) &=& \widetilde{\mathcal{M}}_{gg\rightarrow\gamma}(p,-k,q).
\eea

To find the photon production probability we square each of the matrix elements, sum and average over the initial state and sum over the final state particle polarizations and color. The sum also includes the contribution from the quark flavors. Given the tensor structure of Eq.~(\ref{matrint}), the only polarization involved, both for the gluons and the photon, is the longitudinal one. In this work we account only for the lightest quarks, $f=u,d,s$. Since the processes do not interfere, we sum incoherently the two matrix elements squared, obtaining 
\bea
{\overline{\sum_{\mathrm{\small{c, p,f}}}}}|\widetilde{\mathcal{M}}|^2&=&\mathcal{V}\Delta\tau\dpi^4{\overline{\sum_{\mathrm{\small{c,p,f}}}}}
\nonumber\\
&&
\Biggl[\delta^{(4)}\left(q-k-p\right)|\mathcal{M}_
{gg\rightarrow\gamma}|^2\nonumber\\ 
&+& \delta^{(4)}\left(q+k-p\right)|{\mathcal{M}}_
{g\rightarrow g\gamma}|^2\Biggr],
\label{sumpol1}
\eea
where
\bea
{\overline{\sum_{\mathrm{\small{c,p,f}}}}}|{\mathcal{M}}_
{gg\rightarrow\gamma}|^2&=&{\overline{\sum_{\mathrm{\small{c,p,f}}}}}|{\mathcal{M}}_
{g\rightarrow g\gamma}|^2\nonumber\\
&=&\frac{2\alpha_{\text{em}}\alpha_{\text{s}}^2q_\pp^2}{\pi\omega_q^2}\sum_f q_f^2\left(2\omega_p^2+\omega_k^2+\omega_p\omega_k\right)
\nonumber\\
&\times&
\exp\left\{-\frac{q_\perp^2}{\eB\omega_q^2}\left(\omega_p^2+\omega_k^2+\omega_p\omega_k\right)\right\},\nn\\
\label{sumpol}
\eea
where $N_\text{c}$ is the number of colors. The factor $\mathcal{V}\Delta\tau$ comes from squaring the delta function for energy-momentum conservation in Eq.~(\ref{matrint}). This factor represents the space-time volume where the reaction takes place and consists of the product of the spatial volume of the nuclear overlap region ${\mathcal{V}}(t)$ at time $t$ and the time interval $\Delta\tau$ where the magnetic field can be taken as having a constant intensity $B(t)$. For a given centrality class, the spatial volume can be estimated from the fraction of the number of participants to the total number of nucleons. Since the reaction stops as soon as the magnetic field becomes negligible, the overall lifetime $\Delta{\mathcal{T}}$ can be estimated calculating the duration of the magnetic pulse. To estimate these factors, we perform Monte Carlo simulations using UrQMD~\cite{UrQMD}. This is discussed in the following section.

Notice that in writing Eq.~(\ref{sumpol}) we have already used that, in order to satisfy energy and momentum conservation for massless gluons and photons, and when ignoring the dispersion properties of the magnetized medium, the four-momenta $p^\mu=(\omega_p,\vec{p})$, $k^\mu=(\omega_k,\vec{k})$ and $q^\mu=(\omega_q,\vec{q})$ are related by
\bea
   p^\mu&=&\omega_p(1,\hat{p})
   =\left(\omega_p/\omega_q\right)q^\mu,\nonumber\\
   k^\mu&=&\omega_k(1,\hat{k})
   =\left(\omega_k/\omega_q\right)q^\mu,
\eea
which means that for the reaction to take place, the gluons and the photon are required to have parallel momenta. The invariant photon momentum distribution is given by
\bea
\!\!\!\!\!\!\!\!\!\!\!\!&&\omega_q\frac{dN^{\mbox{\tiny{mag}}}}{d^3q}=\frac{{\mathcal{V}}\Delta \tau}{2(2\pi)^3}
\int\frac{d^3p}{\dpi^32\omega_p}\int\frac{d^3k}{\dpi^32\omega_k}\nonumber\\
\!\!\!\!\!\!&\times&\dpi^4\Bigg{\{}\delta^{(4)}\left(q-k-p\right)n(\omega_p)n(\omega_k){\overline{\sum_{\mathrm{\small{c,p,f}}}}}|{\mathcal{M}}_
{gg\rightarrow\gamma}|^2 \nonumber \\
&& +\delta^{(4)}\left(q+k-p\right)n(\omega_p)\left[1+n(\omega_k)\right]{\overline{\sum_{\mathrm{\small{c,p,f}}}}}|{\mathcal{M}}_
{g\rightarrow g\gamma}|^2\Bigg{\}}, \nonumber \\
\label{invdist}
\eea
where $n(\omega)$ represents the distribution of gluons coming from the shattered glasma. We use for this distribution a simple model that accounts for the high occupation gluon number given by~\cite{Larry2}
\bea
   n(\omega)=\frac{\eta}{e^{\omega/\Lambda_s}-1},
   \label{gluondist}
\eea
where $\eta$ represents the high gluon occupation factor. Notice that this distribution is {\it Bose-Einstein-like}. Thus, the initial state gluon with energy $\omega_k$ comes weighed with an occupation factor  $n(\omega_k)$, whereas the final state one comes weighed with an enhanced occupation factor $1+n(\omega_k)$. We find explicitly

\bea
\frac{1}{2\pi\omega_q}\frac{dN^{\text{mag}}}{d\omega_q}&=&\mathcal{V}\Delta\tau\frac{\alpha_{\text{em}}\alpha_\text{s}^2\pi}{2\dpi^6\omega_q}\sum_f
q_f^2 \nn \\
&\times& 
\int_0^{\omega_q} d\omega_p\left(2\omega_p^2+\omega_q^2-\omega_p\omega_q\right)e^{-g_f(\omega_p,\omega_q)}\nn\\
&\times&
\left\{I_0\left[g_f(\omega_p)\right]-I_1\left[g_f(\omega_p)\right]\right\} \nonumber \\
&\times&
\left\{
n(\wwp)n(|\wq-\wwp|)\right. \nonumber \\ 
&& \left.+n(\wwp)\left[1+n(|\wq-\wwp|)\right]\right\},
\label{yieldexpl}
\eea

where

\bea
g_f(\omega_p,\omega_q)=\frac{\omega_p^2+\omega_q^2-\omega_p\omega_q}{2\eB},
\label{funcg}
\eea
and $I_0$, $I_1$ are the modified Bessel function of the first kind. Notice that $\Delta\tau$ in Eq.~(\ref{yieldexpl}) corresponds to one of the small time intervals that we used to divide the whole time interval $\Delta{\mathcal{T}}$ over which the magnetic pulse is appreciable. The whole yield thus corresponds to the sum of the yields in each of the small time intervals represented by $\Delta\tau$. In each of these intervals the photon emission is considered at the corresponding value of the magnetic field during its time evolution. Therefore, the overall number of photons coming from magnetic field induced processes is not simply proportional to the whole time interval but rather it is computed for each value that the magnetic field takes on during successive small time intervals $\Delta\tau$ and then added up.

In order to compute explicitly the photon distribution and the second harmonic coefficient, recall that the magnitude of the photon momentum transverse (to the direction of the magnetic field), $q_\perp$, is obtained projecting the magnitude of the photon momentum with $\sin (\theta)$, where $\theta$ is the angle between the magnetic field direction and the photon direction of motion. In order to refer $q_\perp$ to the reaction plane, we use that $\sin(\theta) = \sin(\pi/2 - \phi)=\cos(\phi)$, where $\phi$ is the angle between the photon's momentum and the reaction plane direction. The azimuthal distribution with respect to the reaction plane can be given in terms of a Fourier decomposition as
\bea
   \frac{dN^{\mbox{\tiny{mag}}}}{d\phi}=\frac{N^{\mbox{\tiny{mag}}}}{2\pi}
   \left[1+\sum_{n=1}^\infty 2v_n(\omega_q)\cos(n\phi)\right],
\label{Fourier}
\eea
where the total number of photons, $N^{\mbox{\tiny{mag}}}$, is
\bea
N^{\mbox{\tiny{mag}}}=\int d^3q\frac{dN^{\mbox{\tiny{mag}}}}{d^3q}.
\eea

From  Eq.~(\ref{yieldexpl}), and given that $d^3q=\omega_q^2d\omega_qd\phi dy$, where $y$ is the rapidity, it is possible to write, for central rapidity $\Delta y\approx 1$:
\bea
\frac{dN^{\mbox{\tiny{mag}}}}{d\phi}&=&\frac{\alpha_{\text{em}}\alpha_{\text{s}}^2\mathcal{V}\Delta\tau}{2\dpi^5}\cos^2\phi \sum_{f}q_f^2\int_0^{\wq}d\wq'I(\wq',\phi),\nn\\
\label{dNdwqdphi}
\eea
where
\bea
I(\wq',\phi)&=&\int_0^{\omega_q'}d\wwp\left(2\wwp^2+\wq'^2-\wwp\wq'\right)
\nonumber\\
&\times&
e^{-2\cos^2(\phi)g_f(\omega_p,\omega_q')}\nn\\
&\times&\Big\{
n(\wwp)n(|\wq'-\wwp|)
\nn\\
&+& n(\wwp)\left[1+n(|\wq'-\wwp|)\right]\Big\}.
\label{Iwq}
\eea

To find the expression for $v_2$, we take $n=2$ in Eq.~(\ref{Fourier}). Using the orthogonality of the cosine functions of integer multiples of the azimuthal angle, together with the identity
\bea
\int_0^{2\pi}d\phi
&&\!\!\!\!\!\!\cos^2(\phi)\cos(2\phi) e^{-A(x)\cos^2(\phi)}=\pi e^{-A(x)/2}\nn\\
&\times&\left[I_0\left(\frac{A(x)}{2}\right)-\frac{2+A(x)}{A(x)}I_1\left(\frac{A(x)}{2}\right)\right],
\eea
we find from Eq.~(\ref{dNdwqdphi}) that $v_2(\omega_q)$ is given by

\bea
v_2^{\text{mag}}(\omega_q)&=&\frac{\alpha_{\text{em}}\alpha_\text{s}^2\pi\mathcal{V}\Delta\tau}{2\dpi^5 N^{\mbox{\tiny{mag}}} }\sum_f
q_f^2\int_0^{\omega_q} d\omega_q^\prime \int_0^{{\omega_q}^\prime} d\omega_p \nn \\
&\times& 
\left(2\omega_p^2+{\omega^\prime_q}^2-\omega_p\omega_q^\prime \right)e^{-g_f(\omega_p,\omega_q^\prime)}\nn\\
&\times&
\left\{I_0\left[g_f(\omega_p)\right]-\left[1+\frac{1}{g_f(\omega_p)}\right]I_1\left[g_f(\omega_p)\right]\right\} \nn \\
&\times&
\left\{
n(\wwp)n(|\wq^\prime-\wwp|) + n(\wwp)\left[1+n(|\wq^\prime -\wwp|)\right]\right\}. \nn \\
&&
\label{v2expl}
\eea

Notice that the magnetic field dependence of the yield and $v_2^{\text{mag}}$ in Eqs.~(\ref{yieldexpl}) and (\ref{v2expl}), respectively, comes through the function $g_f$ defined in Eq.~(\ref{funcg}). In order to have an estimate for the strength of this magnetic field in the interaction region born out of a Monte Carlo calculation, we use UrQMD. As we proceed to show, the simulation can also be used to estimate  ${\mathcal{V}}$ and $\Delta\mathcal{T}$ (that serves as a proxy for the interaction region lifetime).

\section{Magnetic Field Strength and Space-Time Volume from UrQMD}\label{III}

We use UrQMD~\cite{UrQMD} to compute the time evolution of both, the magnetic field strength and the space-time volume of the interaction region in heavy-ion collisions for a given centrality class. This enables us to perform a systematic study of the photon excess coming from gluon fusion and splitting with a better estimate of the space-time volume (${\mathcal{V}\Delta\mathcal{T}}$) and magnetic field time profile which is an improvement on our previous analysis~\cite{our-old-v2}.

We first simulate  relativistic Au+Au collisions at $\sqrt{s_\text{NN}}=200$ GeV for three centrality classes, 0-20\%, 20-40\% and 40-60\%. Next, for Cu+Cu collisions at the same collision energy, we simulate events in the centrality class 0-40\%. These centrality classes were mapped from impact parameter ranges using the procedure described in Ref.~\cite{CentralityMap}.

\begin{figure}[t!]
    \includegraphics[width=0.5\textwidth]{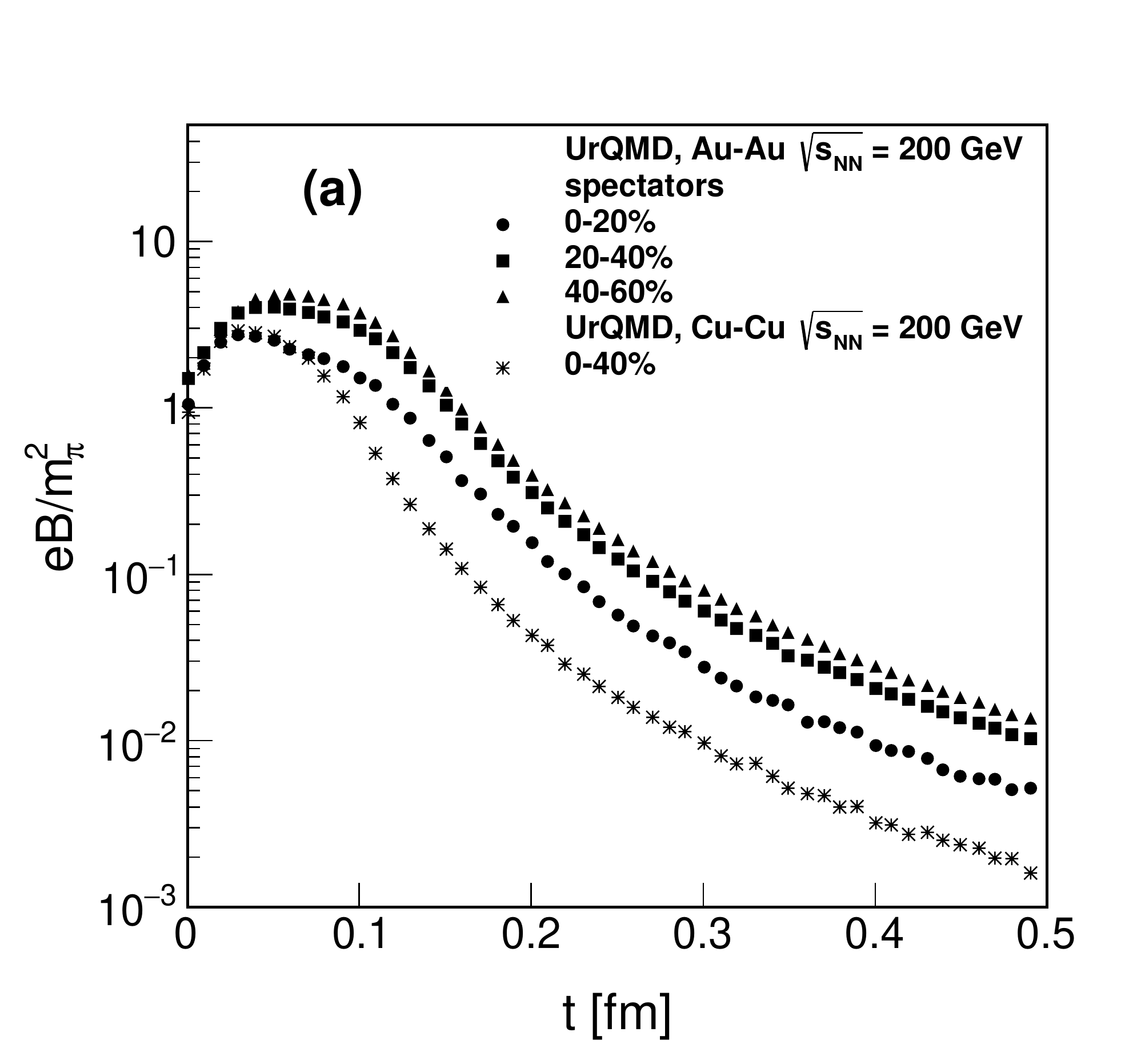}
    \includegraphics[width=0.5\textwidth]{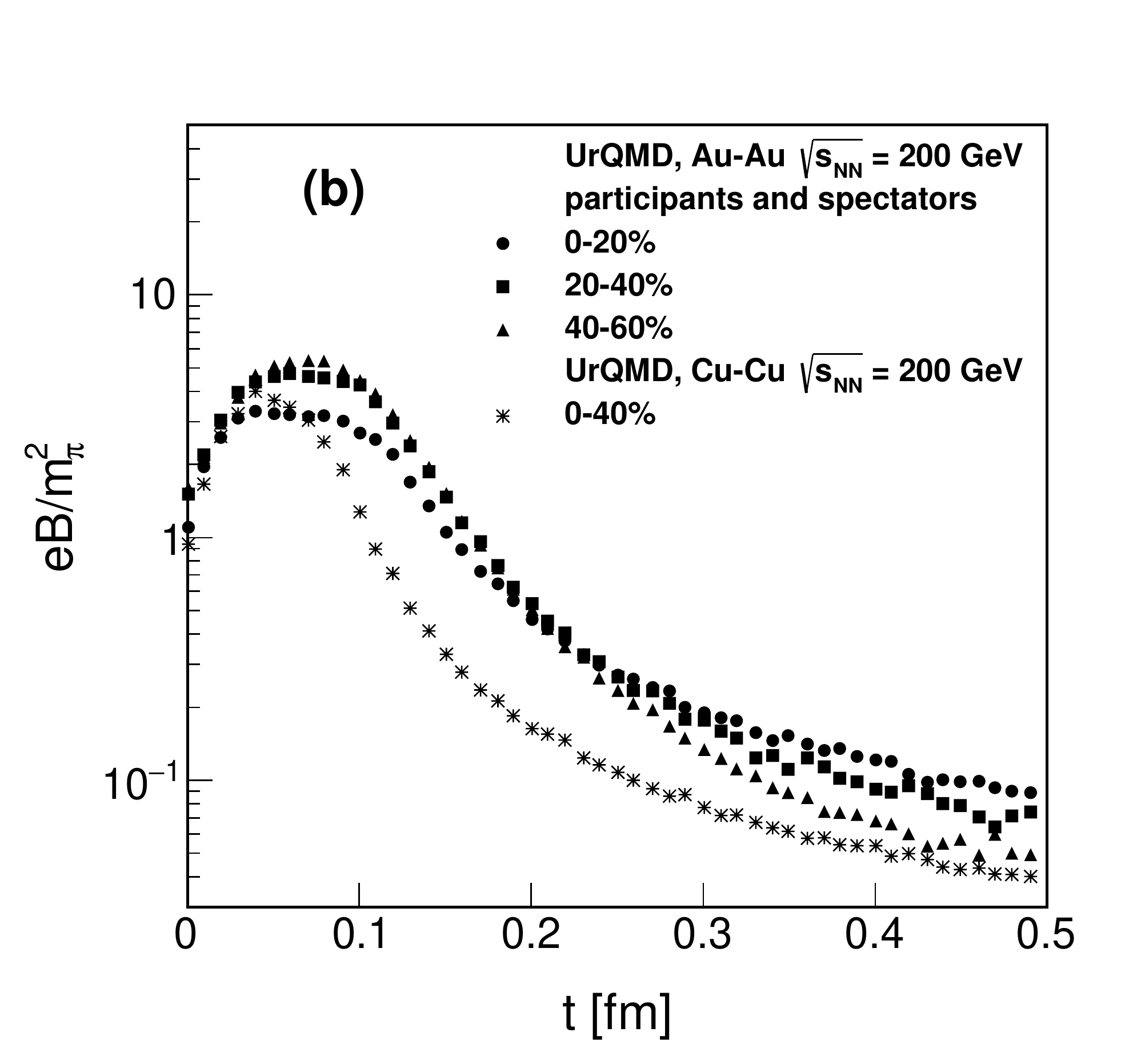}
    \caption{\label{fig:Bfield} Mean magnetic field strength produced by (a) spectators and (b) spectators and participants at the middle of the interaction region as a function of time for three centrality classes 0-20\%, 20-40\% and 40-60\% in Au+Au collisions and one centrality class 0-40\% in Cu+Cu collisions at
    $\sqrt{s_\text{NN}}=200$ GeV.}
\end{figure}

For each event, the magnetic field at position $\mathbf{x}$ and at time $t$ can be calculated using  the Liénard-Wiechert potential generated by non-accelerated charges moving along the beam direction~\cite{LWpotential} as
\bea
e\mathbf{B}(\mathbf{x},t) = \alpha_{em} \sum_{j}  
\frac{(1-v_j^{2})~\mathbf{v}_j\times \mathbf{R}_j}{ R_j^{3}\left[1-\frac{(\mathbf{v}_i\times\mathbf{R}_j)^{2}}{R_j^{2}}\right]^{3/2}},
\label{magneticfield}
\eea
where $\mathbf{R}_j = \mathbf{x} - \mathbf{x}_j(t)$, $\mathbf{x}_j(t)$ is the position of the $j$-th charge moving with velocity $\mathbf{v}_j$, $R_j$ is the magnitude of $\mathbf{R}_j$ and the sum runs over charged particles in each event. The magnetic field is computed at the middle of the interaction region, $\mathbf{x}=0$. Figure~\ref{fig:Bfield} (a) [(b)] shows the time profile of this magnetic field at the center of the collision region produced by spectators (participants and spectators), for different centrality ranges in Au+Au and Cu+Cu collisions at $\sqrt{  s_{\text{NN}} }=200$ GeV. The overall time interval was taken to be $\Delta{\mathcal{T}} = t-t_0 $ with $t_0=0$ fm being the beginning of the collision. Since the magnetic pulse is short-lived, this time interval was set to $\Delta{\mathcal{T}}=0.5$ fm with time varying in steps of $\Delta\tau=0.01$ fm. Although  hydrodynamical simulations start at early times of order 0.2-0.6 fm~\cite{hydro-photons1,Chatterjee,Shen}, notice that the strength of the magnetic field has decreased by more than an order of magnitude from its initial value by times of order 0.2 fm. This means that possible thermal effects, implied by hydrodynamical  simulations, start on average after the time interval when the magnetic pulse is important. Also, notice that at the earliest times, although the Cu+Cu system is smaller than the Au+Au one, the considered centrality class produces a similar average field strength for the former compared to the latter for the most central collisions.

\begin{figure}[t!]
    \includegraphics[width=0.5\textwidth]{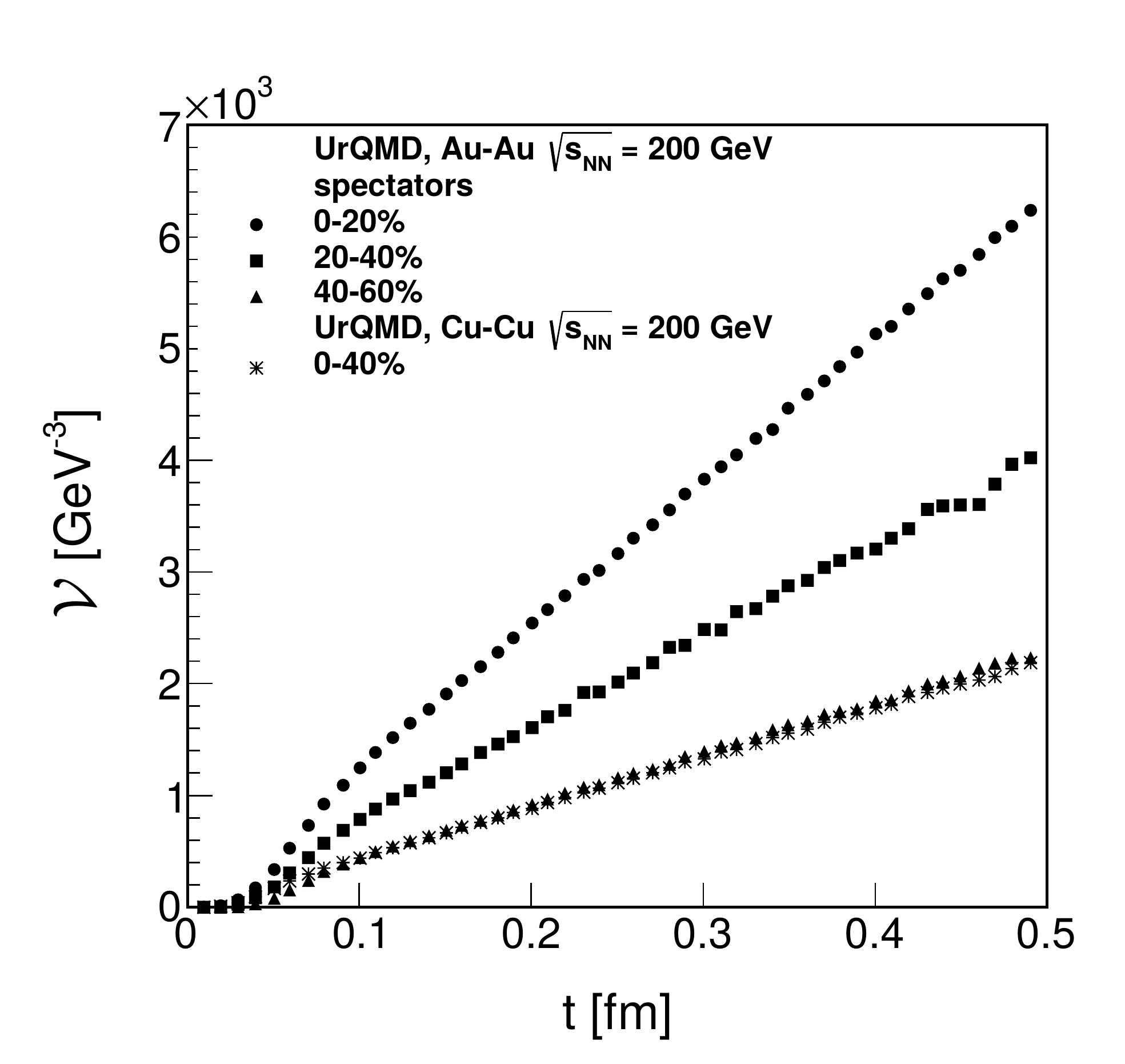}
    \caption{ Volume ${\mathcal{V}}$ as a function of time for three centrality classes 0-20\%, 20-40\% and 40-60\% in Au+Au collisions and one centrality class 0-40\% in Cu+Cu collisions at $\sqrt{s_\text{NN}}=200$ GeV.}
    \label{fig:Volume}
\end{figure}

Figure~\ref{fig:Volume} shows the volume ${\mathcal{V}}(t)$ calculated for each centrality class as
\begin{equation}
\mathcal{V}(t) = 2 t\pi r_\text{A}^2 \left(\frac{N_\text{part}}{2N}\right)^{2/3},
\label{volume}
\end{equation}
\begin{figure}[t!]
   \includegraphics[width=0.5\textwidth]{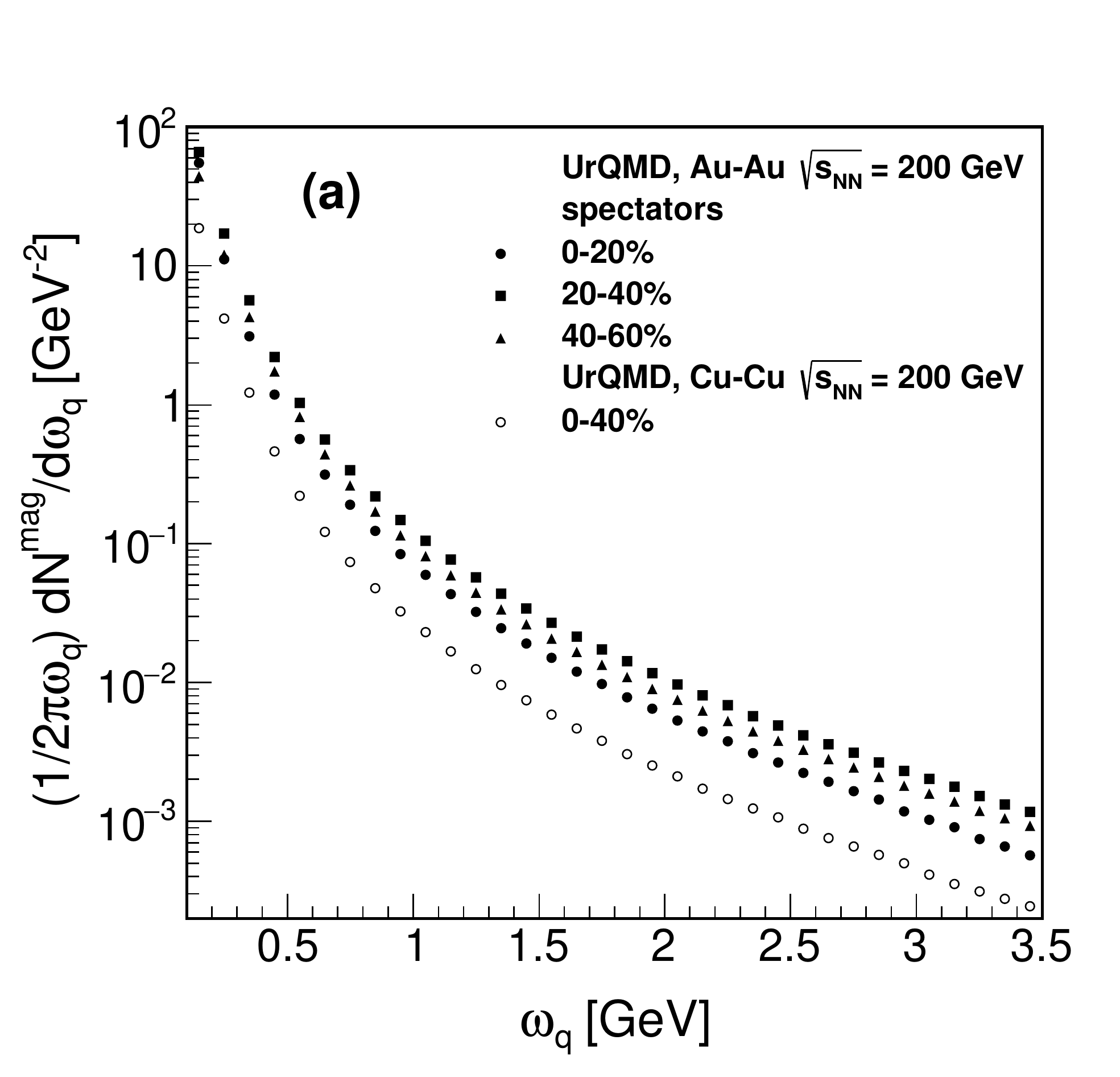}
   \includegraphics[width=0.5\textwidth]{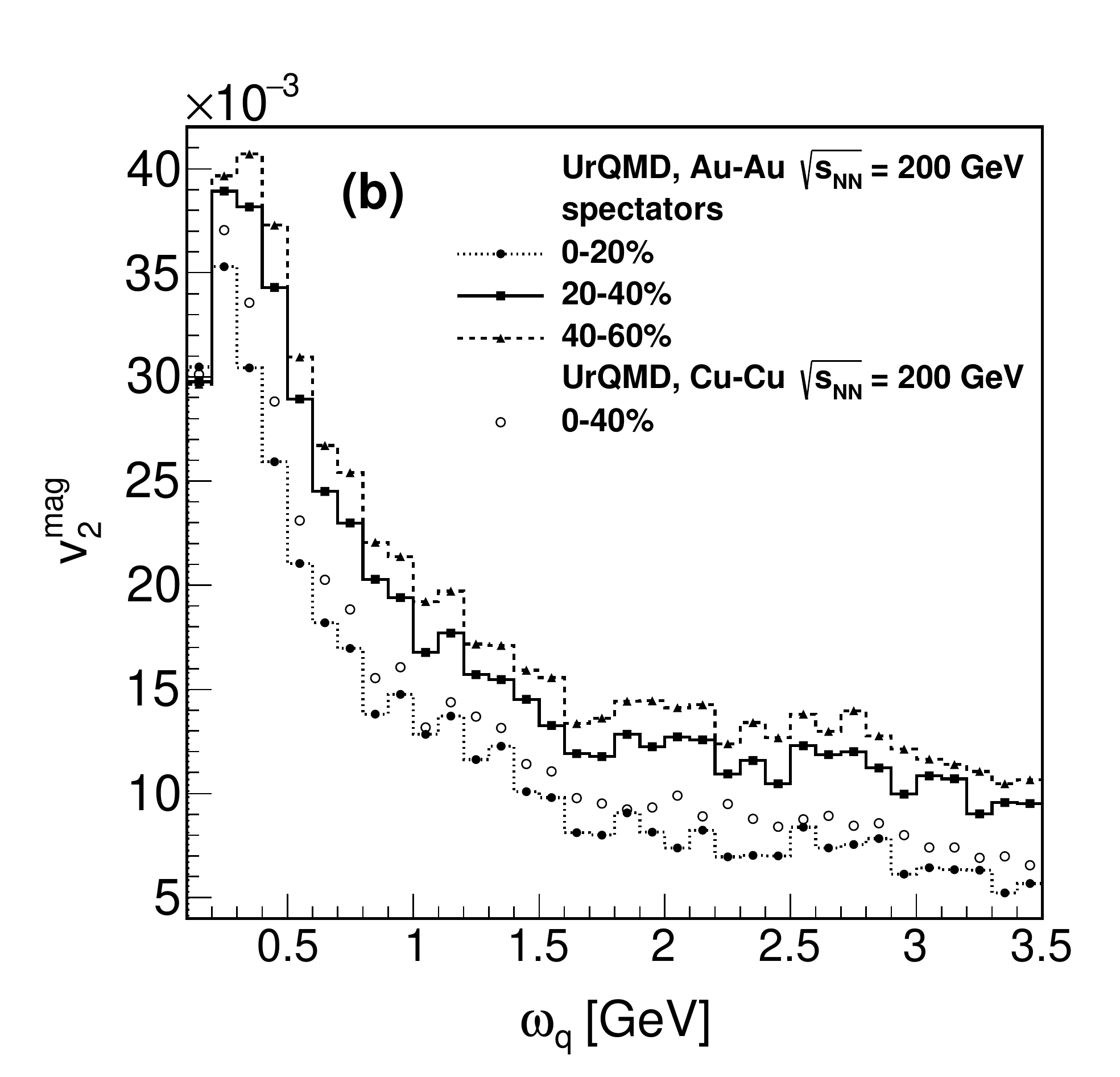}
    \caption{
    (a) Invariant momentum distribution
    and (b) $v_2^{\text{mag}}$ 
    for Au+Au collisions in the 0-20\%, 20-40\% and 40-60\% centrality classes and Cu+Cu collisions in the 0-40\% centrality class at $\sqrt{  s_{\text{NN}} }=200$ GeV considering only the magnetic field generated by spectators.} \label{fig5}
\end{figure}

\begin{figure}[t!]
    \centering
   \includegraphics[width=0.5\textwidth]{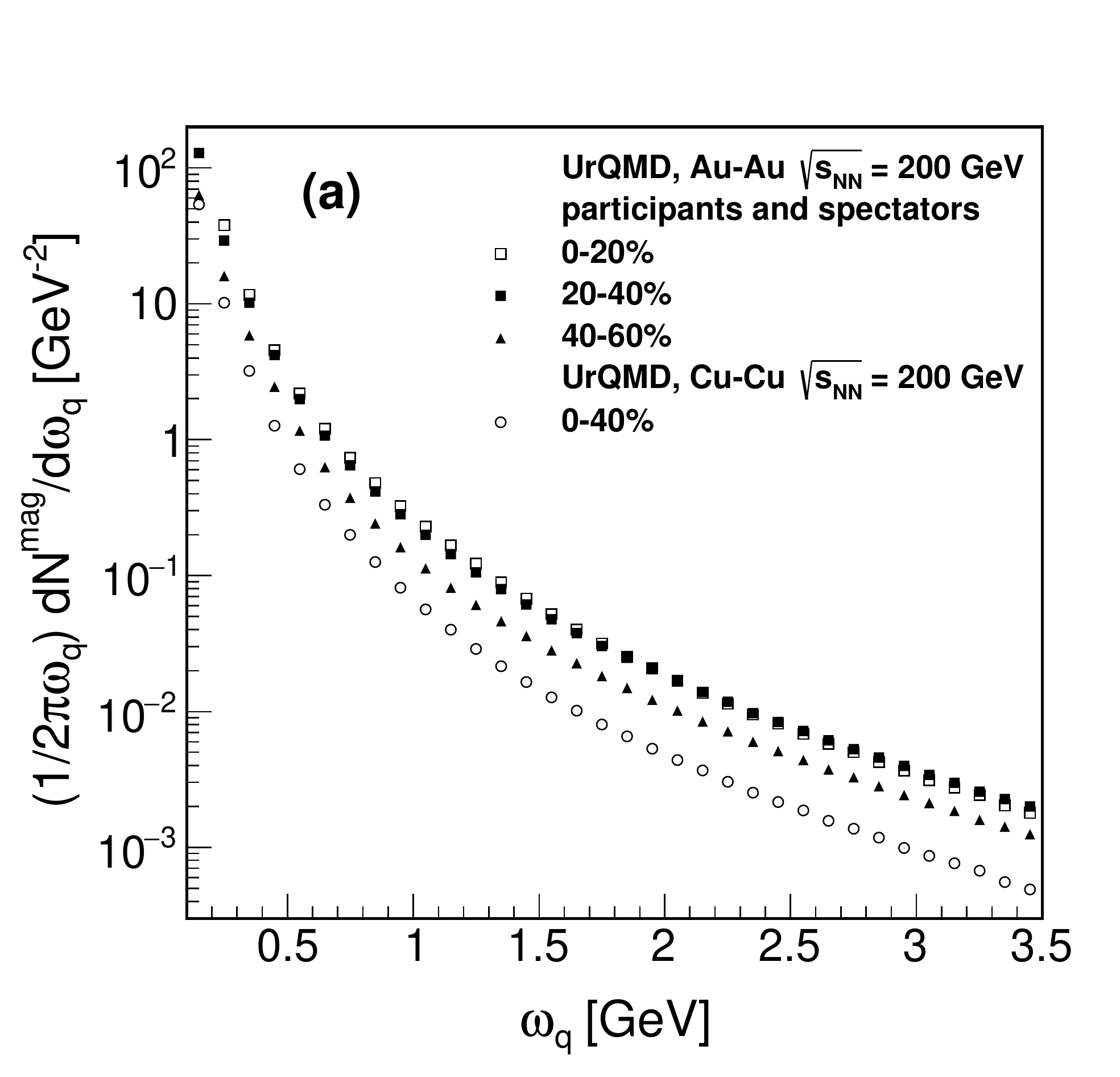}
   \includegraphics[width=0.5\textwidth]{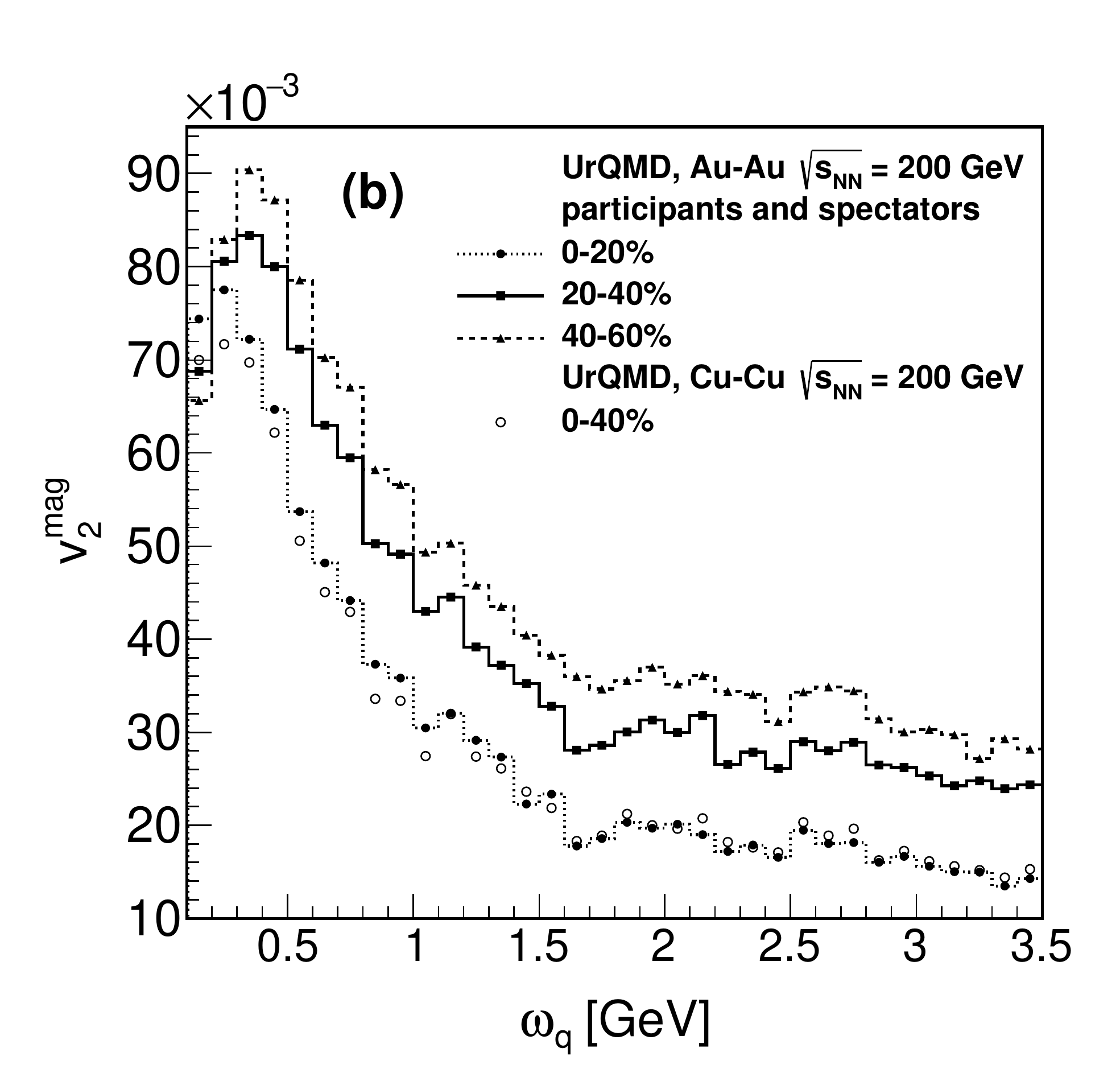}
   \caption{
    (a) Invariant momentum distribution
    and (b) $v_2^{\text{mag}}$ 
    for Au+Au collisions in the 0-20\%, 20-40\% and 40-60\% centrality classes and Cu+Cu collisions in the 0-40\% centrality class at $\sqrt{  s_{\text{NN}} }=200$ GeV considering the magnetic field generated by both, the participants and spectators.}
\label{fig6}
\end{figure}
\begin{figure}[t]
    \centering
   \includegraphics[width=0.5\textwidth]{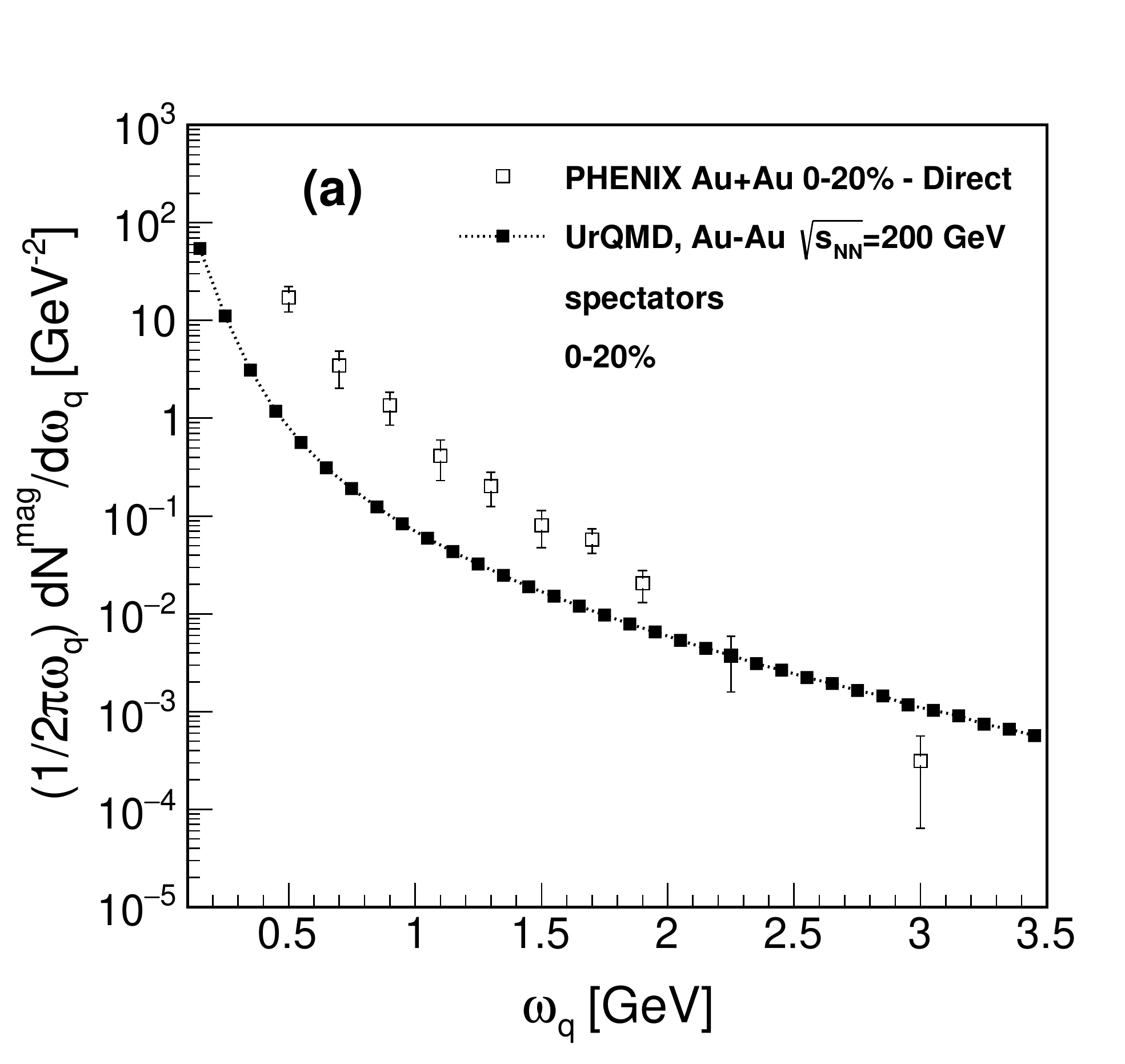}
   \includegraphics[width=0.5\textwidth]{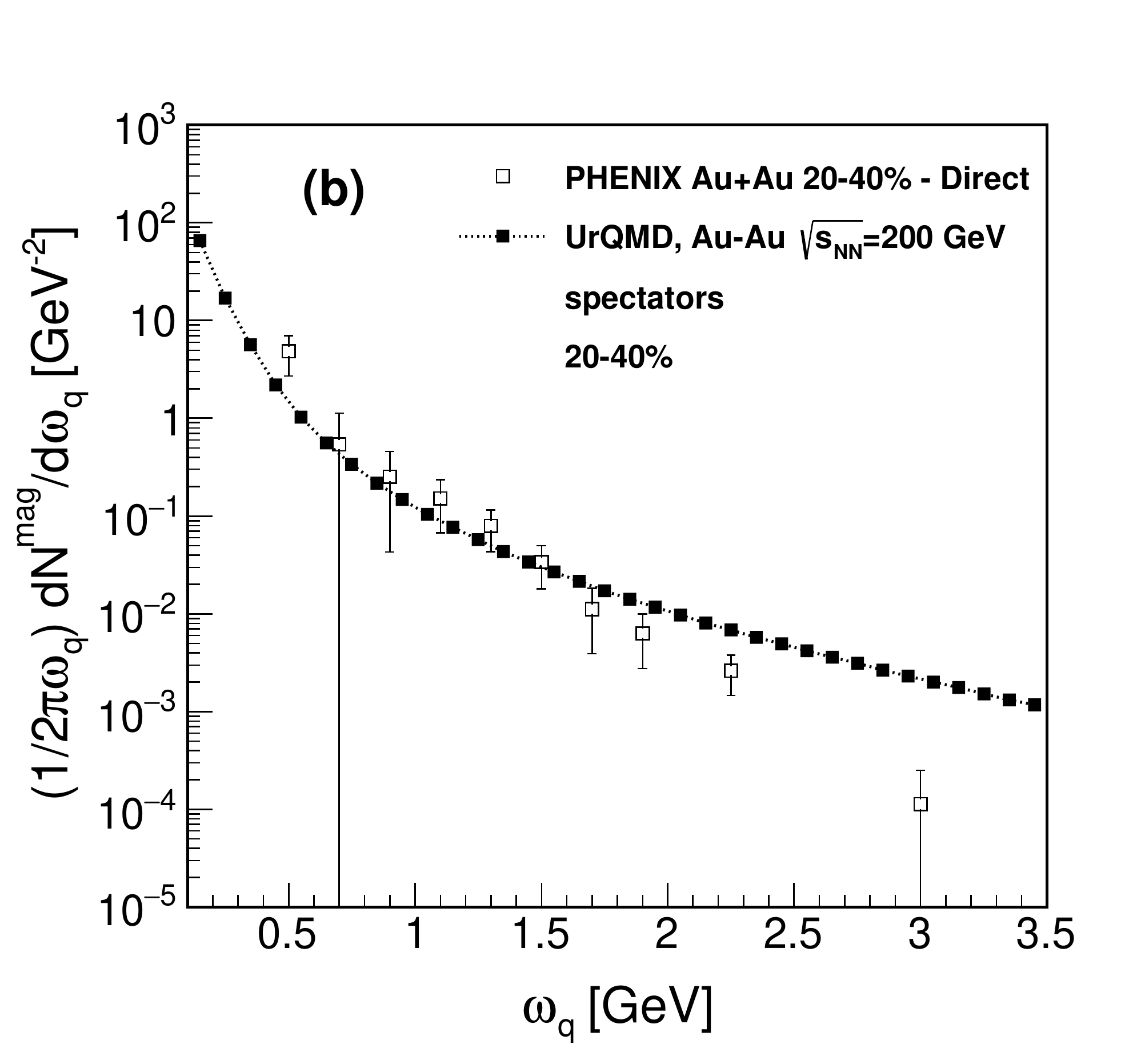}
   \caption{Difference between PHENIX data~\cite{experimentsyield} and the hydrodynamical calculation of Ref.~\cite{hydro-photons1} (open symbols) for Au+Au collisions at $\sqrt{s_{NN}}=200$ GeV for (a) the 0-20\% and (b) 20-40\% centrality classes, compared to the calculation of the invariant yield (filled symbols) considering only the contribution to the magnetic field strength produced by the spectators.}
    \label{fig7}
\end{figure}
\begin{figure}[t!]
    \centering
   \includegraphics[width=0.5\textwidth]{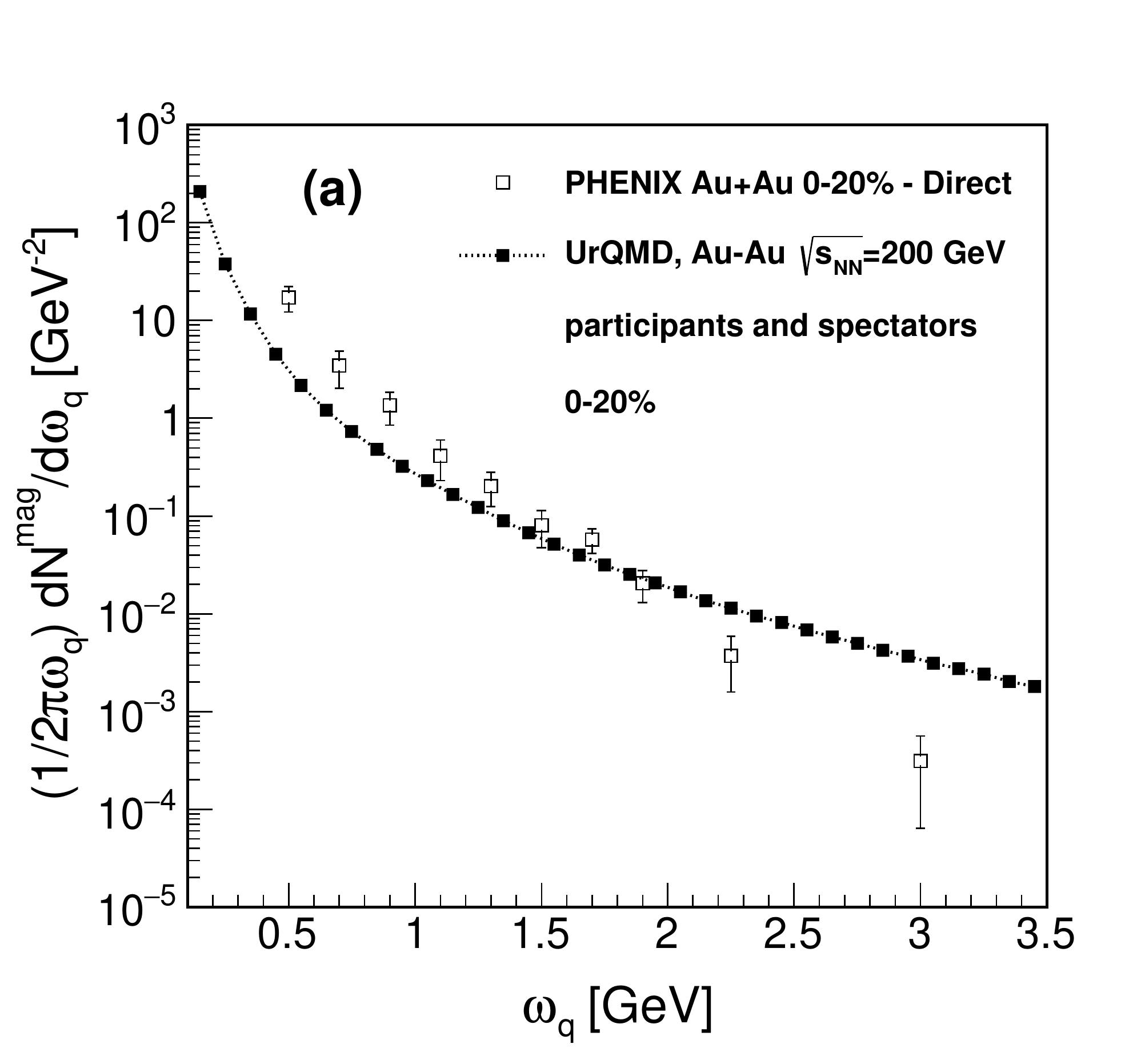}
   \includegraphics[width=0.5\textwidth]{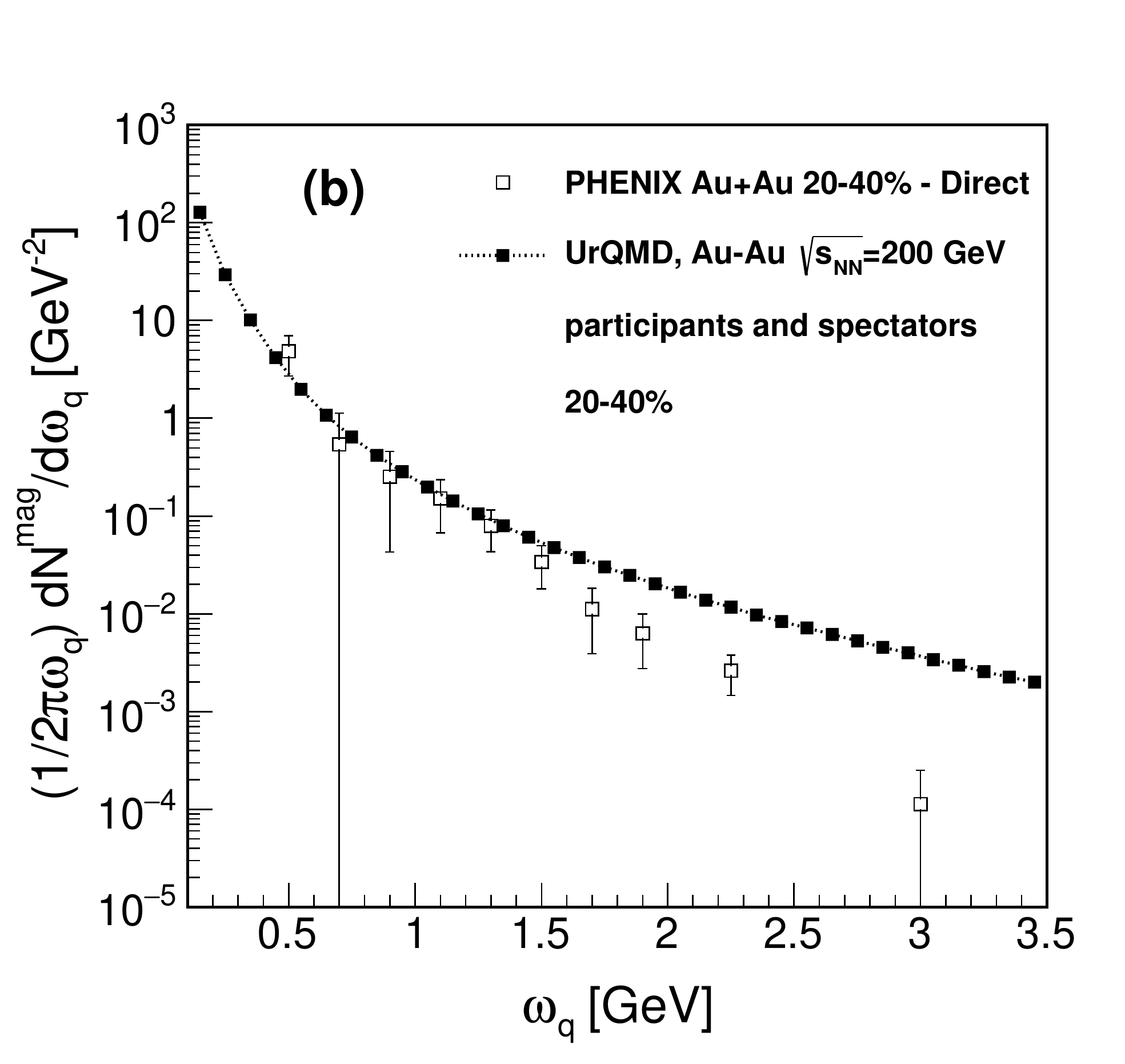} 
    \caption{Difference between PHENIX data~\cite{experimentsyield} and the hydrodynamical calculation of Ref.~\cite{hydro-photons1} (open symbols), for Au+Au collisions at $\sqrt{s_{NN}}=200$ GeV for (a) the 0-20\% and (b) 20-40\% centrality classes, compared to the calculation of the invariant yield (filled symbols) considering the contribution to the magnetic field strength produced by both the participants and spectators. }
    \label{fig8}
\end{figure}
where $\pi r_\text{A}^2 (N_\text{part}/2N)^{2/3}$ corresponds to the transverse area of a cylinder having a radius $r_\text{A}$ ($r_\text{Au} = 6.38$ fm, $r_\text{Cu} = 3.9$ fm~\cite{AuRadius}), weighed with the ratio of the number of participants $N_\text{part}$ to twice the number of nucleons $2N$ in each of the colliding nuclei, raised to the power $2/3$. This last factor accounts for the overlap area in the collision region. The cylinder symmetry axis is taken to be along the beam. The factor $2t$ accounts for the longitudinal distance between the receding nuclei moving close to the speed of light after time $t$.  

\section{Results}\label{IV}

\begin{figure}[t!]
    \centering
   \includegraphics[width=0.5\textwidth]{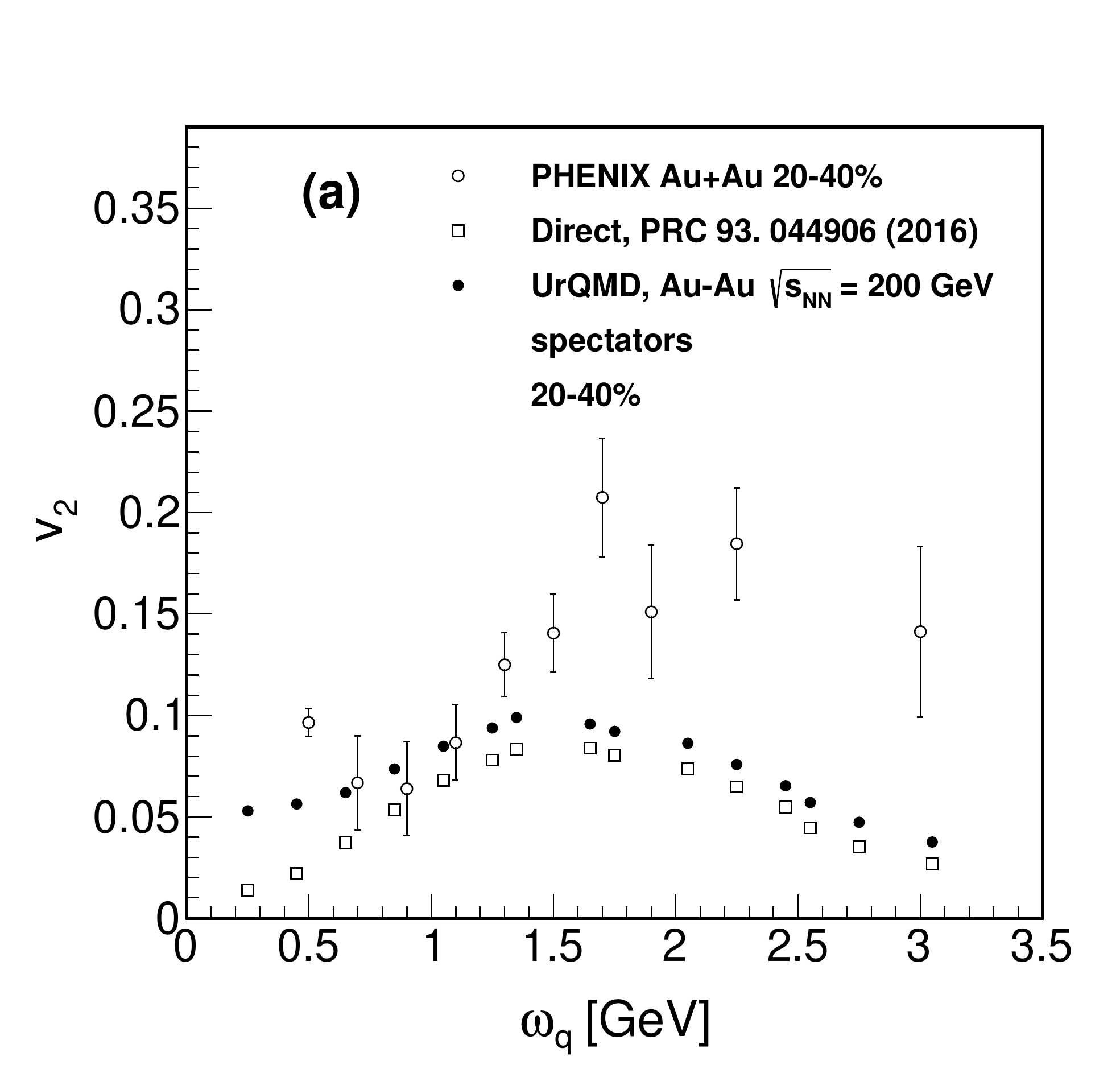}
   \includegraphics[width=0.5\textwidth]{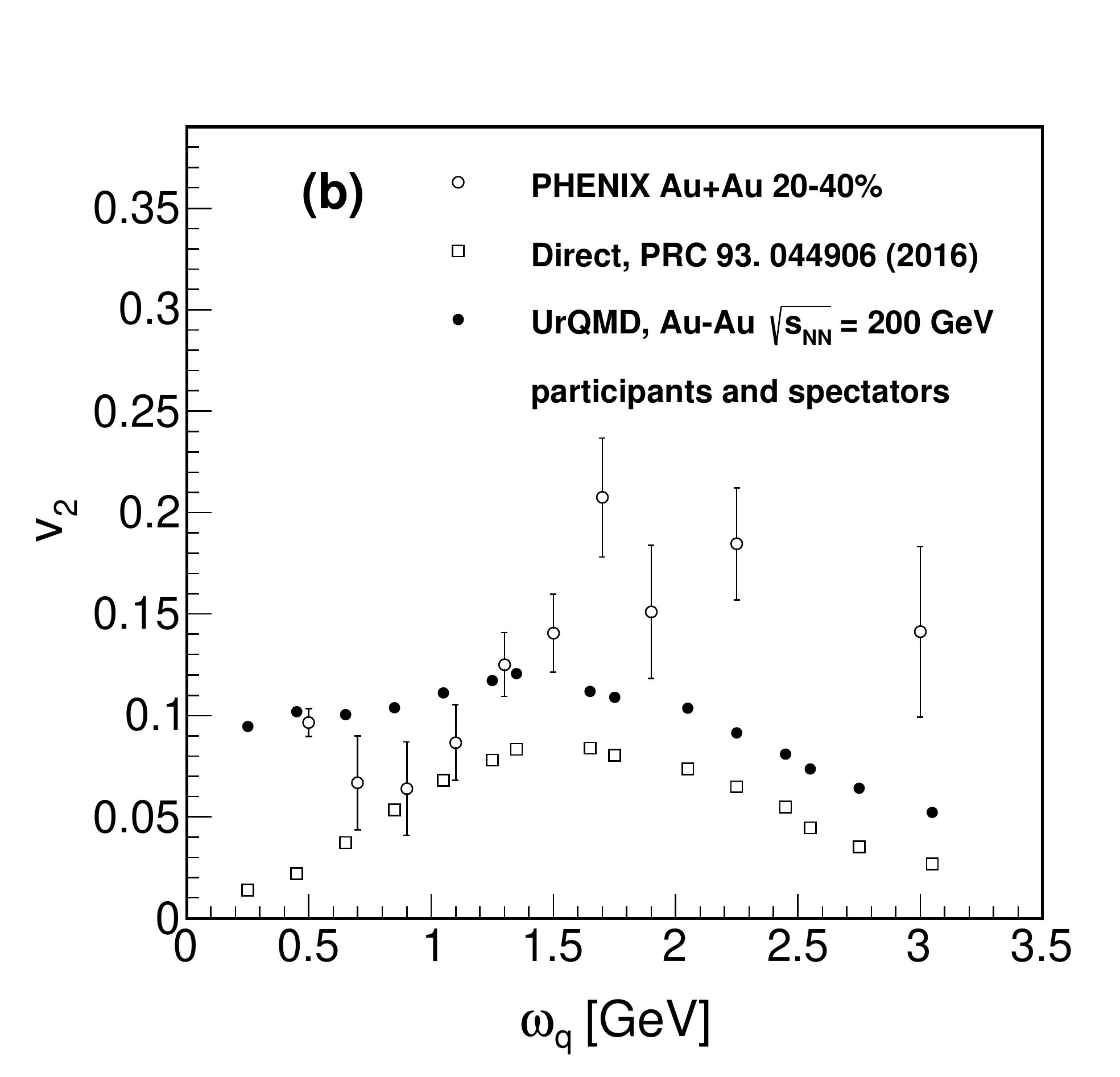} \caption{$v_2$ as a function of the photon energy $\omega_q$. Shown are the contributions from the hydrodynamical calculation of Ref.~\cite{hydro-photons1} (open squares) and the magnetic field-dependent calculation summed as a weighed average including the field strength coming from (a) the spectators and (b) the spectators and participants (filled circles). The experimental data (open symbols) correspond to Au+Au collisions at $\sqrt{s_{NN}}=200$ GeV in the 20-40\% centrality class from Ref.~\cite{experimentsyield}.}
    \label{fig9}
\end{figure}
Throughout this section we present results using $\alpha_s=0.3 $, $\Lambda_s=2 $ GeV, $\eta=3$. Figure~\ref{fig5} shows the invariant momentum distribution and corresponding $v_2^{\text{mag}}$ , for Au+Au collisions at $\sqrt{  s_{\text{NN}} }=200$ GeV in the 0-20\%, 20-40\% and 40-60\% centrality ranges and Cu+Cu collisions in the 0-40\% centrality range at the same energy. We use the magnetic field strength generated only by spectators. Notice that the net yield is calculated by adding the yields for all time intervals within $\Delta\mathcal{T}$ and is given by
\bea
\frac{dN^{\text{mag}}}{d\omega_q} &=& \sum_{i=1} \left[\frac{dN^{\text{mag}}}{d\omega_q}\right]_i,
\eea
where $\left[\frac{dN^{\text{mag}}}{d\omega_q}\right]_i$ is the yield corresponding to the $ith$-time interval $\Delta\tau_i$, given by Eq.~(\ref{yieldexpl}).
The corresponding net $v_2$ is computed as a weighted average which accounts for the time-varying yield, as
\bea
v_2^{\text{mag}}(\omega_q) &=& \frac{
\sum_{i=1} \left[
\frac{dN^{\mbox{\tiny{mag}}}}{d\omega_q}(\omega_q)
\right]_i 
[v_2^{\text{mag}}(\omega_q)]_i
}{\sum_{i=1} \left[\frac{dN^{\mbox{\tiny{mag}}}}{d\omega_q}\right]_i}
\eea
where $[v_2^{\text{mag}}(\omega_q)]_i$ is the harmonic coefficient corresponding to the $ith$-time interval $\Delta\tau_i$, given by Eq.~(\ref{v2expl}). 
Figure~6 shows the invariant momentum distribution and corresponding $v_2^{\text{mag}}$
for the same systems as in Figure~\ref{fig5} but with a magnetic field strength  generated by both spectators and participants. Notice that including the contribution to the magnetic field strength coming from the spectators and participants, produces an increase of both the invariant yield and $[v_2^{\text{mag}}(\omega_q)]$.

We now proceed to compare these results to experimental data. Notice that since the magnetic contribution represents an excess over calculations not including these effects, for the yield we can compare to the difference between data and a hydrodynamical simulation of the direct photon yield.  However, for $v_2$, one needs to consider a weighed average between the $v_2$ contribution from photons produced by the magnetic field and the $v_2$ contribution from the direct photons produced by the simulation. For this comparison we choose the state-of-the-art calculation of Ref.~\cite{hydro-photons1}.


Figure~\ref{fig7}~(\ref{fig8}) shows the difference between PHENIX data~\cite{experimentsyield} and the hydrodynamical calculation of Ref.~\cite{hydro-photons1} --open symbols-- compared to our calculation of the invariant yield --filled symbols-- considering only the contribution to the magnetic field strength produced by the spectators (spectators and participants),
in Au+Au collisions at $\sqrt{  s_{\text{NN}} }=200$ GeV. (a) shows the 0-20\% and (b) the 20-40\% centrality classes. Notice that when considering the magnetic field produced only by spectators the theoretical yield compares better to peripheral than to central collisions. The comparison improves both for peripheral and central collisions when the contribution to the field strength from the participants is also included. Notice that the photon yield coming from magnetic field induced process is normalized to the total number of photons produced by this process during the whole time interval where the magnetic field is appreciable. In this sense, this yield is already a weighted sum, since it takes into account adding the number of photons emitted during each of the small time intervals, which in turn are already normalized to the number of photons in each of these intervals.

Figure~\ref{fig9} shows $v_2$ as a weighted average accounting for the magnetic and direct photons
\bea
  && v_2(\omega_q)=\nn\\
&&\frac{
\sum_{i=1}^m \left[
\frac{dN}{d\omega_q}
\right]_i 
[v_2^{\text{mag}}(\omega_q)]_i
   +
   \frac{dN^{\mbox{\tiny{direct}}}}{d\omega_q}(\omega_q)\
   v_2^{\mbox{\tiny{direct}}}(\omega_q)} 
   {\sum_{i=1}^m \left[
\frac{dN}{d\omega_q}
\right]_i 
   + 
   \frac{dN^{\mbox{\tiny{direct}}}}{d\omega_q}(\omega_q)},\nonumber\\
   \label{v2weighted}
\eea
where $dN^{\mbox{\tiny{direct}}}/d\omega_q$ and $v_2^{\mbox{\tiny{direct}}}$ are the ($\omega_q$-dependent) spectrum and second harmonic coefficient of direct photons from Ref.~\cite{hydro-photons1}, respectively, compered to the experimental data for Au+Au collisions at $\sqrt{s_{NN}}=200$ GeV in the 20-40\% centrality class from Ref.~\cite{experimentsyield}. (a) corresponds to the calculation of magnetic field effects coming only from the spectators the spectators and (b) from the the spectators and participants. Notice that the magnetic field contribution improves the agreement with experimental data for the low part of the spectrum helping to describe the rise of $v_2$ as the photon energy decreases.

As can be seen from Figs.~\ref{fig8} and~\ref{fig9}, the excess photon yield and $v_2$ coming from magnetic field induced gluon fusion and splitting helps to better describe the experimental data having as a baseline a state-of-the-art calculation accounting for many of the well described sources of photons. The effect on the photon yield is to increase the distribution and at the same time shift it to higher photon energy values. For the case of $v_2$, the agreement of the calculation with data is particularly good in the lowest part of the spectrum since it describes well the observed experimental fall between 0.5 and 1 GeV. This fall has received little attention and in our approach it is due to the rise and fall of the $v_2$ distribution that peaks for energy values of the order of $\sqrt{eB}$. For the energy region above 1 GeV the calculation overshoots the data. This may be due to the fact that the obtained photon distribution we used contains a power-like tail that overestimates the yield for large photon energies. This is a shortcoming of our approach that considers $|eB|$ to be the largest energy scale in the problem and which is bound to fail for large photon energies. An improved matrix element that replaces Eq.~(\ref{matrint}), valid for large photon energies is being computed and will be reported elsewhere.

\section{Summary and conclusions}\label{V}

We have computed the contribution to the photon yield and $v_2$ from gluon fusion and splitting induced by a magnetic field during the early stages of a relativistic heavy-ion collision, where there is a large gluon occupation number below the saturation scale $\Lambda$. Although $\Lambda\gg |eB|$, notcie that nowhere in the calculation it is necessary to directly compare $|eB|$ to $\Lambda$. Thus, the calculation describes the situation where the magnetic field driven photon emission happens in the background of a large amount of gluons whose momentum is smaller than $\Lambda$ and whose occupation number decreases (exponentially) fast as their momentum approaches $\Lambda$. Nevertheless this occupation number is still appreciable for $|eB| < \omega_q^2 < \Lambda^2$ and the calculation is thus only valid  for low photon energies $\omega_q < \Lambda$. For larger photon energies, the approximation breaks down and one needs to resort to another approximation to describe the time evolution of emitted gluons in this kinematic regime. In this sense, our treatment has to be regarded as a first analytical step towards an estimation of the photon yield obtained from a magnetic field driven process.

The magnetic field strength and volume are computed using UrQMD simulations and the results compared with recent data from PHENIX. For the yield, the excess coming from the magnetic field induced processes is compared with the difference between PHENIX data and the hydrodynamical calculation of Ref.~\cite{hydro-photons1}. $v_2$ is computed as a weighed average accounting for magnetic and direct photons. The results show a relatively good agreement for the lower part of the spectra and is better for peripheral collisions. The comparison improves when the magnetic field strength includes the contribution from both spectators and participants. 
We emphasize that our results point into the direction of enhancing the yield and $v_2$ for $p_T\lesssim 1$ GeV/c, a trend not incompatible with data~\cite{Gabor}. However, uncertainties on experimental data are still too large and analyses with larger data sets are eagerly awaited for.

There are several avenues for improvement. For example, we can consider a spatial dependence of $\Lambda_s$ and $\eta$ that we use in the simple model that accounts for the high occupation gluon number given by Eq.~(\ref{gluondist}). Here, in order to capture the essence of the magnetic field effect for photon production, we have considered only a treatment of these parameters as if averaged in space, inspired in Ref.~[17]. This is enough for a short-lived magnetic pulse. Notice also that we calculate the magnetic field strength at ${\bf x}={\bf 0}$. This approximation is enough when considering a pulse with spatial spread of $\Delta x\sim 1/\sqrt{eB}$. Indeed, since for the widest spread scenario, the maximum value for the field intensity is $eB \sim 5 m_\pi^2$, the spatial spread is about $\Delta x \sim 1.5$ fm. This is a narrow spatial spread. For larger magnetic field strengths the spread becomes even narrower. Thus, the approximation where we consider the field strength evaluated at ${\bf x}={\bf 0}$ is justified in the context of this analysis.

Finally, in order to access a better description of the photon yield for larger photon energies, the approximations leading to the used matrix element need to be relaxed. The calculation of an improved matrix element is being currently explored and will be reported elsewhere.


\section*{ACKNOWLEDGEMENTS}

The authors thank L.A. Hern\'andez and S. Hern\'andez-Ortiz for useful discussions. Support for this work has been received in part by UNAM-DGAPA-PAPIIT grant number AG100219 and by Consejo Nacional de Ciencia y Tecnolog\1a grant number 256494.
%
%

\end{document}